\documentclass[english]{emulateapj}

\usepackage{graphicx}
\usepackage{apjfonts}
\def\gs{\mathrel{\raise1.16pt\hbox{$>$}\kern-7.0pt %
\lower3.06pt\hbox{{$\scriptstyle \sim$}}}}         %
\def\ls{\mathrel{\raise1.16pt\hbox{$<$}\kern-7.0pt %
\lower3.06pt\hbox{{$\scriptstyle \sim$}}}}         %

\slugcomment{}
\shorttitle{Spitzer IRS Spectra of 70 micron Sources}
\shortauthors{Brand et al.}

\makeatother
\begin{document}

\title{$Spitzer$ Mid-Infrared Spectroscopy of 70 $\rm \mu m$ Selected Distant Luminous Infrared Galaxies}

\author{Kate Brand\altaffilmark{1}, Dan ~W. Weedman\altaffilmark{2}, Vandana Desai\altaffilmark{3}, Emeric Le Floc'h\altaffilmark{4,5}, Lee Armus\altaffilmark{6}, Arjun Dey\altaffilmark{7}, Jim ~R. Houck\altaffilmark{2}, Buell ~T. Jannuzi\altaffilmark{7}, Howard ~A. Smith\altaffilmark{8}, B.~T. Soifer\altaffilmark{3,6}} 

\altaffiltext{1}{Space Telescope Science Institute, 3700 San Martin Drive, Baltimore, MD 21218; brand@stsci.edu}
\altaffiltext{2}{Astronomy Department, Cornell University, Ithica, NY 14853}
\altaffiltext{3}{Division of Physics, Mathematics and Astronomy, California Institute of Technology, 320-47, Pasadena, CA 91125}
\altaffiltext{4}{Institute for Astronomy, University of Hawaii, 2680 Woodlawn Drive, Honolulu, HI 96822, USA}
\altaffiltext{5}{Spitzer fellow}
\altaffiltext{6}{Spitzer Science Center, California Institute of Technology, 220-6, Pasadena, CA 91125}
\altaffiltext{7}{National Optical Astronomy Observatory, 950 North Cherry Avenue, Tucson, AZ 85726} 
\altaffiltext{8}{Harvard-Smithsonian Center for Astrophysics, 60 Garden Street, Cambridge, MA 02138}

\begin{abstract}
We present mid-infrared spectroscopy obtained with the {\it Spitzer Space Telescope} of a sample of 11 optically faint, infrared luminous galaxies selected from a {\it Spitzer} MIPS 70$\rm \mu m$ imaging survey of the NDWFS Bo\"otes field. These are the first $Spitzer$ IRS spectra presented of distant 70$\rm \mu m$-selected sources. All the galaxies lie at redshifts $0.3<z<1.3$ and have very large infrared luminosities of L$_{IR}\sim 0.1-17 \times 10^{12} \rm~L_\odot$. 
Seven of the galaxies exhibit strong emission features attributed to polycyclic aromatic hydrocarbons (PAHs). The average IRS spectrum of these sources is characteristic of classical starburst galaxies, but with much larger infrared luminosities. The PAH luminosities of $\nu$L$_{\nu}$(7.7$\mu$m)$\sim 0.4 - 7 \times 10^{11} \rm~L_\odot$ imply star formation rates of $\sim 40 - 720$ M$_\odot$ yr$^{-1}$. Four of the galaxies show deep 9.7$\mu$m silicate absorption features and no significant PAH emission features (6.2$\rm \mu m$ equivalent widths $<$ 0.03$\rm \mu m$). The large infrared luminosities and low $\nu$f$_{\nu}$(70$\mu$m)/$\nu$f$_{\nu}$(24$\mu$m) flux density ratios suggests that these sources have AGN as the dominant origin of their large mid-infrared luminosities, although deeply embedded but luminous starbursts cannot be ruled out. If the absorbed sources are AGN-dominated, a significant fraction of all far-infrared bright, optically faint sources may be dominated by AGN.
\end{abstract}


\section{Introduction}

Luminous and ultra-luminous infrared galaxies (LIRGs: L(8-1000$\rm \mu m$)$>$ 10$^{11}$ L$_\odot$; \citealt{san88}) have been studied extensively in the local Universe with the Infrared Astronomical Telescope (IRAS; \citealt{soi87}; \citealt{san90}), the Infrared Space Observatory (ISO; e.g., \citealt{lut98}; \citealt{gen00}; \citealt{tra01}), and more recently, with the Infrared Spectrograph (IRS; \citealt{hou04}) on $Spitzer$ (\citealt{wee05}; \citealt{brl06}; \citealt{arm07}; \citealt{des07}). These galaxies exhibit a large range of properties in the mid-IR, some showing strong PAH emission features characteristic of powerful (up to $\approx$1000 M$_\odot$ yr$^{-1}$) star formation rates (e.g., \citealt{brl06}; \citealt{smi07}), and all exhibiting a large range in 9.7$ \rm \mu m$ silicate absorption or emission strengths (e.g., \citealt{wee05}; \citealt{des07}; \citealt{ima07}). $Spitzer$ IRS is now enabling the study of the mid-infrared spectra of LIRGs to much higher redshifts ($z\sim$2.6; \citealt{hou05} \citealt{yan05}). Although rare locally, LIRGs become an important population at high redshifts and account for an increasing fraction of the star-formation activity in the Universe \citep{lef05}. By studying their infrared properties, one can estimate the extent to which AGN and star-formation contribute to their infrared luminosities, and therefore determine a correct census of starbursts and AGN at epochs in the Universe when their luminosity density was at its maximum.

A particularly interesting population of high redshift optically faint infrared sources has been discovered using the Multiband Imaging Photometer (MIPS; \citealt{rie04}). Various observing programs with $Spitzer$ IRS have found that MIPS sources at flux density levels of f$_{\nu}$(24$\rm \mu m$) $\sim$ 1 mJy with optical magnitudes $R$ $\ga$ 24 Vega magnitudes are typically at z $\sim$ 2 (\citealt{hou05}; \citealt{yan07}; \citealt{wee06}). Most of the sources are characterized by strong absorption by the 9.7$\rm \mu m$ silicate feature, but sources chosen with an additional indicator of star formation (sub-mm detection or shape of the spectral energy distribution) often shown strong polycyclic aromatic hydrocarbon (PAH) emission features (\citealt{lut05}; \citealt{wee06b}; \citealt{men07}; \citealt{yan07}; \citealt{saj07}). Using traditional optical techniques to characterize these sources is difficult because they are too faint in the optical band. Mid-infrared spectroscopy is the best currently available tool in understanding these sources but we need to learn how to categorize them from the different features that are exhibited in the infrared band. We might expect sources which are selected on the basis of their large far-infrared flux densities to contain large amounts of cool dust and show signatures of starbursts. Mid-infrared spectroscopy of these sources will test this assumption and may help our understanding of high redshift LIRGs as a whole.

In this paper, we present the first {\it Spitzer} IRS spectra of a small sample of galaxies selected on the basis of their large 70$\rm \mu m$ flux densities and optical faintness. We have previously reported results on sources chosen only with criteria of large infrared to optical flux ratios, as measured by comparison of their 24 $\rm \mu m$ flux density, f$_{\nu}$(24$\rm \mu m$) with optical magnitude ($R$-[24]>14; \citealt{hou05}; \citealt{wee06}). We now extend the IRS observations to additional sources in Bo\"{o}tes that are less extreme in their $R$-[24] colors but are selected with the additional criteria of detection at 70$\mu$m by $Spitzer$ MIPS. 

A cosmology of $H_0 = 70 {\rm ~km ~s^{-1} ~Mpc^{-1}}$, $\Omega_M$=0.3, and $\Omega_\Lambda$=0.7 is assumed throughout. 

\section{Source Selection}

We selected a sample of optically faint far-infrared luminous sources from the {\it Spitzer} MIPS 70$\rm \mu m$ survey of the NOAO Deep Wide-Field (NDWFS; \citealt{jan99}) Bo\"otes field for follow-up with $Spitzer$ IRS. Such sources have not previously been studied and our paper provides the first census of their basic mid-IR properties. The 70$\rm \mu m$ observations reach a 5$\sigma$ limiting flux density of f$_{70\rm \mu m}$=25 mJy, and yield a total of $\approx$ 330 sources. To insure a highly reliable 70$\rm \mu m$ catalog, and allow follow-up observations with IRS, we only included 70$\rm \mu m$ sources with a 24$\rm \mu m$ flux density, f$_{24}>$1 mJy and f$_{70\rm \mu m}>$30 mJy (this resulted in only 16 rejected sources and should not bias our sample to sources with unusually high 24$\rm \mu m$ flux densities). We measured optical photometry in the publicly available NDWFS $R$-band images, and selected all sources with $R\ge$20~Vega mag. Although ideally, we would like to select sources as red in their optical to infrared colors as the $R-$[24]$>$14 sources previously observed in the Bo\"otes field, the lack of available candidates has required us to relax this criteria. Our sample have a larger range of optical to infrared colors corresponding to 11$<R$-[24]$<$16.5. Although the optical obscuration is likely to be less extreme, the combined faint optical and bright infrared selection should still select distant infrared bright galaxies. The basic optical and infrared properties of the 11 sources are presented in Table~\ref{tab:70sample}. The IRAC photometry is from the IRAC Shallow Survey \citep{eis04}.

\begin{deluxetable*}{llllllllllllll}
\tabletypesize{\scriptsize}
\setlength{\tabcolsep}{0.05in}
\tablecolumns{14} 
\tablewidth{0pc}
\tablecaption{\label{tab:70sample} Optical and IR properties.} 
\tablehead{ 
\colhead{IRS ID} & \colhead{MIPS} & \colhead{$B_W$$^{a,b}$} & \colhead{$R$$^{a,b}$} & \colhead{$I$$^{a,b}$} & \colhead{$K$$^{a,b}$} & \colhead{f$_{3.6}$$^c$} & \colhead{f$_{4.5}$$^c$} & \colhead{f$_{5.8}$$^c$} & \colhead{f$_{8}$$^c$} & \colhead{f$_{24}$$^c$} & \colhead{f$_{70}$} & \colhead{f$_{160}$} &\colhead{R-[24]}\\
\colhead{} & \colhead{name} & \colhead{mag} & \colhead{mag} & \colhead{mag} & \colhead{mag} & \colhead{mJy} & \colhead{mJy} & \colhead{mJy} & \colhead{mJy} & \colhead{mJy} & \colhead{mJy} & \colhead{mJy} & \colhead{mag}\\
}
\startdata
70Bootes1   &SST24 J142651.9+343135   & 22.7  &20.6  &19.8  &16.7$^e$& 0.11 &  0.09 & 0.08 &  0.17  &  1.39& 32.8$\pm$4.2&$<$120& 11.3\\
70Bootes2   &SST24 J142732.9+324542   & 22.7  &20.7  &19.9  &$-$     & 0.07 &  0.07 & 0.09 &  0.21  &  1.22& 34.6$\pm$4.1&$<$120& 11.2\\
70Bootes3   &SST24 J143639.0+345222   & 23.2  &22.0  &21.1  &$-$     & 0.13 &  0.10 & 0.10 &  0.12  &  1.26& 35.0$\pm$6.3&145$\pm$29& 12.6\\
70Bootes4   &SST24 J143218.1+341300   & 23.3  &22.0  &21.0  &17.8$^d$& 0.10 &  0.09 & 0.09 &  0.12  &  1.22& 36.6$\pm$5.1&100$\pm$20& 12.6\\
70Bootes5   &SST24 J143050.8+344848   & $-$   &24.6  &22.4  &$-$     & 0.04 &  0.04 & 0.08 &  0.24  &  4.25& 43.2$\pm$5.1&70$\pm$14& 16.5\\
70Bootes6   &SST24 J143830.6+344412   & 23.8  &21.1  &18.7  &$-$     & 0.13 &  0.14 & 0.22 &  0.78  &  3.19& 45.2$\pm$4.0&$<$120& 12.7\\
70Bootes7   &SST24 J143151.8+324327   & 22.3  &20.5  &19.7  &$-$     & 0.20 &  0.16 & 0.17 &  0.37  &  2.18& 51.8$\pm$4.7&135$\pm$27& 11.7\\ 
70Bootes8   &SST24 J143341.9+330136   & 22.7  &21.1  &20.2  &17.8$^d$& 0.08 &  0.06 & 0.09 &  0.21  &  4.72& 63.2$\pm$3.9&105$\pm$21& 13.1\\   
70Bootes9   &SST24 J143820.7+340233   & 22.7  &20.3  &19.3  &15.7$^d$& 0.40 &  0.27 & 0.32 &  0.34  &  3.46& 67.2$\pm$3.1&245$\pm$49& 12.0\\
70Bootes10  &SST24 J143449.3+341014   & 23.6  &21.5  &20.7  &17.5$^d$& 0.25 &  0.23 & 0.68 &  0.97  &  2.19& 94.5$\pm$4.5&120$\pm$24& 12.7\\
70Bootes11$^f$&SST24 J143205.6+325835 & 21.3  &20.1  &19.8  &$-$     & 0.08 &  0.13 & 0.27 &  1.3   & 16.66&115.8$\pm$12.8&$<$120& 13.5\\
\enddata
\tablenotetext{a}{All quoted magnitudes are Vega magnitudes from the NDWFS DR3 (Jannuzi et al. in preparation).}
\tablenotetext{b}{Errors on $B_w$, $R$, $I$, and $K$-band magnitudes are $<$0.1.}
\tablenotetext{c}{Errors on f$_{3.6}$, f$_{4.5}$, f$_{5.8}$, f$_{8}$, and f$_{24}$ flux densities are $<$0.1 mJy.}
\tablenotetext{d}{$K_s$-mag from FLAMEX survey \citep{els06}.}
\tablenotetext{e}{$K$-mag from NDWFS survey Dey et al. in preparation.}
\tablenotetext{f}{Object 8 in Houck et al.~2007.}
\end{deluxetable*} 
\section{Observations and Data Reduction}

The spectroscopic observations were made with the IRS Short Low module in order 1 only (SL1) and with the Long Low module in orders 1 and 2 (LL1 and LL2), described in \citet{hou04}.  These orders give low resolution spectral coverage from $\sim$8\,$\mu$m to $\sim$35\,$\mu$m.  Sources were placed on the slits by offsetting from a nearby 2MASS star. The integration times for individual sources are given in Table~\ref{tab:irs}.

Because these faint sources are dominated by background signal, we restrict the number of pixels used to define the source spectrum, applying an average extraction width of only 4 pixels (which scales with wavelength).  This improves the signal-to-noise ratio, although some source flux in outlying pixels is lost so a correction is needed to change the fluxes obtained with the narrow extraction to the fluxes that would be measured with a standard extraction.  This flux correction is derived empirically by extracting an unresolved source of high S/N with both techniques and is a correction of about 10\%, although the correction varies with order and with wavelength. 

The background which was subtracted for LL1 or LL2 includes co-added backgrounds from both nod positions having the source in the other slit (i.e., both nods on the LL1 slit when the source is in the LL2 slit), added together with the alternative nod position in the same slit, yields a background observation with three times the integration time as for the source.  For SL1, there was no separate background observation with the source in the SL2 slit, so background subtraction was done between co-added images of the two nod positions in SL1.  Independent spectral extractions for each nod position were compared to reject highly outlying pixels in either spectrum, and a final mean spectrum was produced.  Data were processed with version 13.0 of the SSC pipeline, and extraction of source spectra was done with the SMART analysis package \citep{hig04}. 
\begin{deluxetable*}{llllllllllll}
\tabletypesize{\scriptsize}
\setlength{\tabcolsep}{0.05in}
\tablecolumns{12}
\tablewidth{0pc}
\tablecaption{\label{tab:irs} IRS properties and derived physical properties.} 
\tablehead{ 
\colhead{IRS} & \colhead{exp. time$^a$} & \colhead{IRS} & \colhead{$\rm\nu L_{\nu}(6\mu m)$} & \colhead{$\rm\nu L_{\nu}(7.7\mu m)$$^b$} & L$_{IR}$(7.7)$^c$ & L$_{IR}$(24+70+160)$^d$ & SFR$^e$ &\colhead{f(15)/f(6)} & \colhead{S$_{10}$$^f$} & \colhead{PAH EW$^g$} & \colhead{IRS} \\
\colhead{ID} & \colhead{s} & \colhead{$z$} & \colhead{log($\rm erg~s^{-1}$)} & \colhead{log($\rm erg~s^{-1}$)} & \colhead{log(L$_\odot$)} & \colhead{log(L$_\odot$)} & \colhead{M$_\odot$ yr$^{-1}$} & \colhead{} & \colhead{} & \colhead{$\rm \mu m$} & \colhead{class$^h$}  \\
}
\startdata
70Bootes1 &480,1200 & 0.501 &  43.89 & 44.53 & 11.72 &12.15 & 90   & 4.2 & $-$ & 0.26$\pm$0.05&sb\\
70Bootes2 &480,1200 & 0.366 &  43.65 & 44.18 & 11.36 &11.79 & 40   & 1.8 & $-$ & 0.85$\pm$0.17&sb\\
70Bootes3 &600,1440 & 0.986 &  44.69 & 45.37 & 12.56 &13.05 & 625  & 4.3 & $-$ & 0.48$\pm$0.05&sb\\
70Bootes4 &480,1200 & 0.975 &  44.24 & 45.32 & 12.51 &12.95 & 565  & 9.6 & $-$ & 0.59$\pm$0.07&sb\\
70Bootes5 &240,480  & 1.21  &  45.51 & $-$   & $-$   &13.24 & $-$  & 7.1 & -1.0& $<$0.02 &abs\\
70Bootes6 &240,480  & 0.94  &  45.27 & $-$   & $-$   &13.01 & $-$  & 4.0 & -1.9& $<$0.01 &abs\\
70Bootes7 &480,960  & 0.664 &  44.46 & 45.20 & 12.39 &12.61 & 430  & 3.4 & $-$ & 0.29$\pm$0.04& sb\\ 
70Bootes8 &240,480  & 0.81  &  44.79 & $-$   & $-$   &12.88 & $-$  & 7.7 & -3.6& $<$0.03 &abs\\   
70Bootes9 &240,480  & 0.668 &  44.62 & 45.43 & 12.62 &12.81 & 720  & 4.0 & $-$ & 0.53$\pm$0.04&sb\\
70Bootes10&480,960  & 0.512 &  44.43 & 44.80 & 11.98 &12.41 & 170  & 2.1 & $-$ & 0.13$\pm$0.01&sb\\
70Bootes11$^i$&240,480& 0.48&  44.75 & $-$   & $-$   &12.49 & $-$  & 8.3 & -2.2& $<$0.02 & abs\\
\enddata
\tablenotetext{a}{Total integration times in SL1 (first entry) and each of LL1 and LL2 (second entry; same for each).}
\tablenotetext{b}{7.7$\rm \mu m$ luminosity in source rest frame, determined from peak flux density of 7.7$\rm \mu m$ feature without continuum subtraction.}
\tablenotetext{c}{Infrared luminosity derived using the relation: log[L$_{IR}$]=log[$\rm\nu L_{\nu}(7.7\mu m)$] + 0.78 from Houck et al.~(2007).}
\tablenotetext{d}{Minimum infrared luminosity estimated from L$_{IR}$=[$\rm\nu L_{\nu}(24\mu m)$+$\rm\nu L_{\nu}(70\mu m)$+$\rm\nu L_{\nu}(160\mu m)$]$\times$(1+$z$).}
\tablenotetext{e}{Star formation rate derived from infrared luminosity using relation from \citet{ken98}.}
\tablenotetext{f}{Silicate absorption strength as defined in Section. 4.2.}
\tablenotetext{g}{Rest-frame equivalent width of 6.2$\rm \mu m$ PAH emission feature.}
\tablenotetext{h}{PAH-dominated and absorption-dominated sources are denoted ``sb'' and ``abs'' respectively. See Section 4 for definitions of these terms.}
\tablenotetext{i}{Object 8 in Houck et al.~2007.}
\end{deluxetable*} 
\section{IRS Spectral Characteristics}

\begin{figure*}[h]
\begin{center}
\includegraphics[height=37mm]{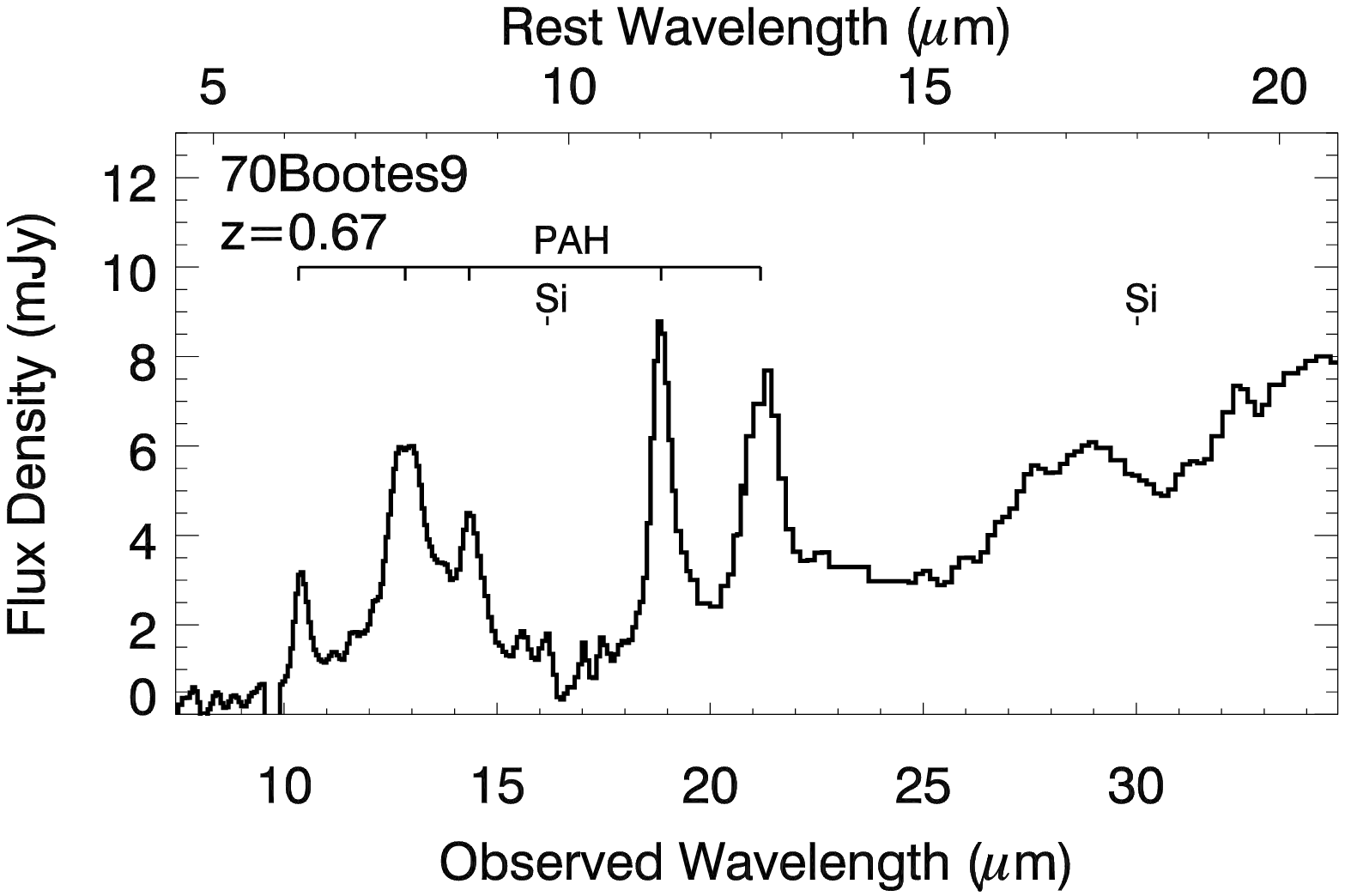}
\includegraphics[height=37mm]{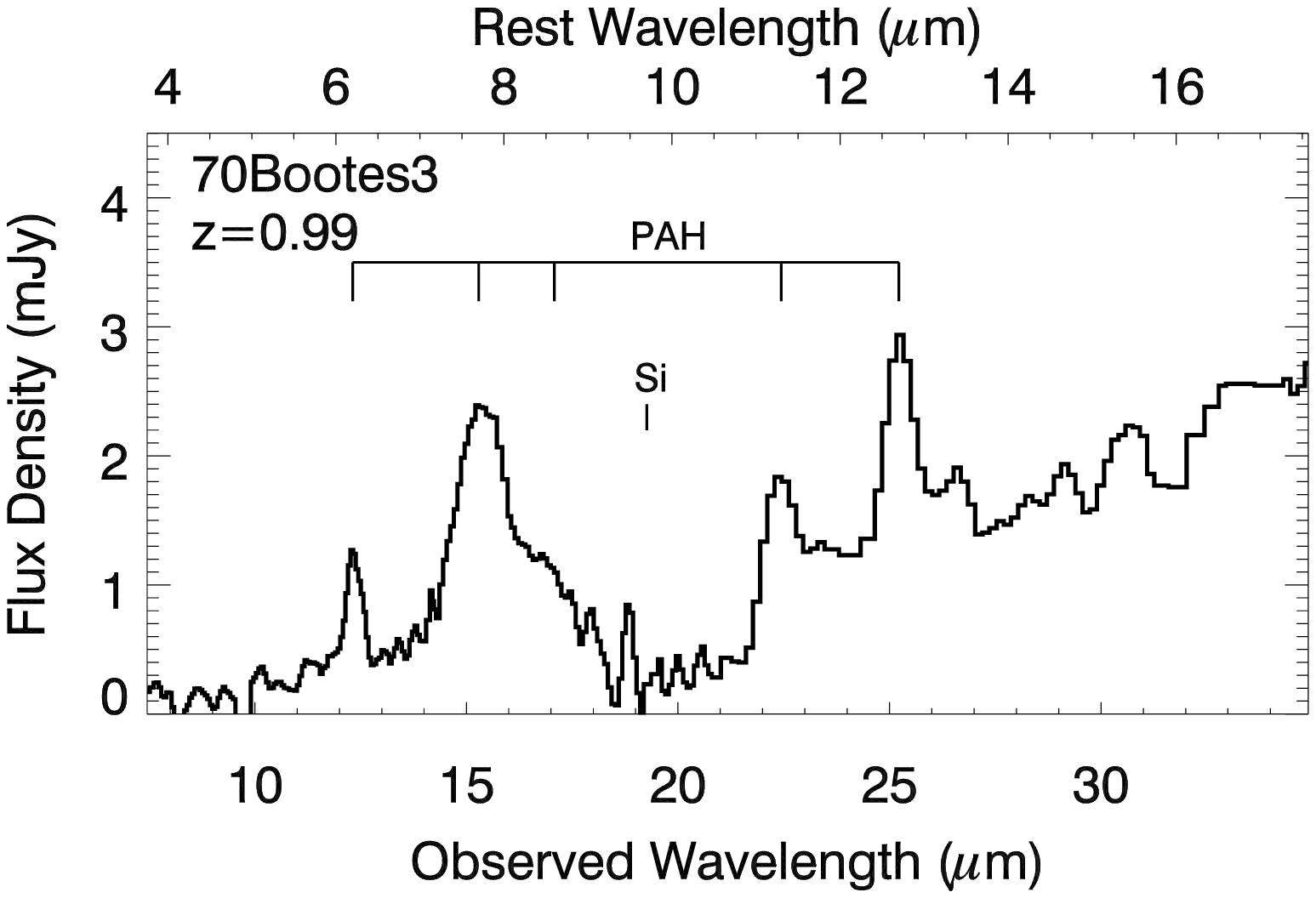}
\includegraphics[height=37mm]{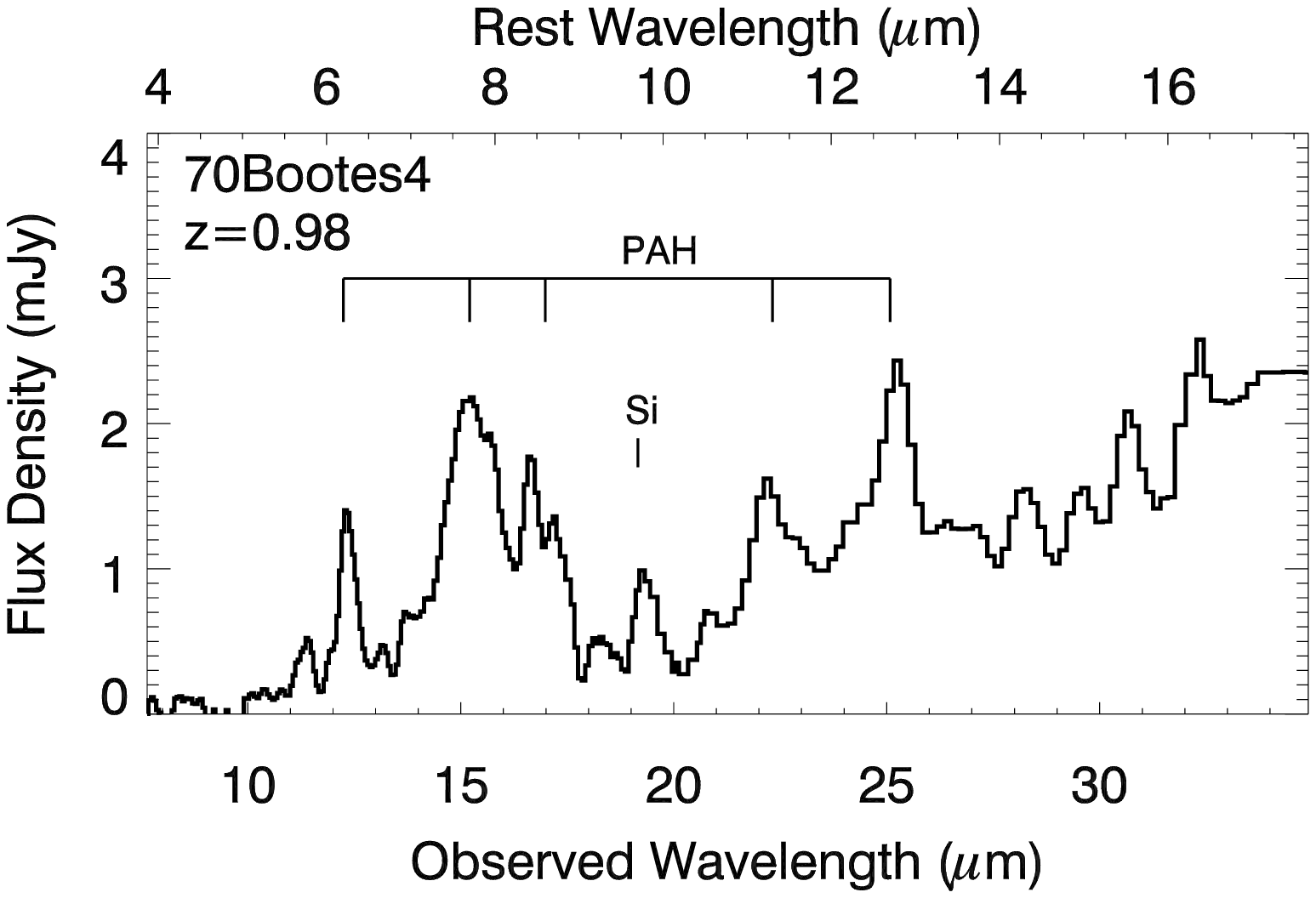}
\includegraphics[height=37mm]{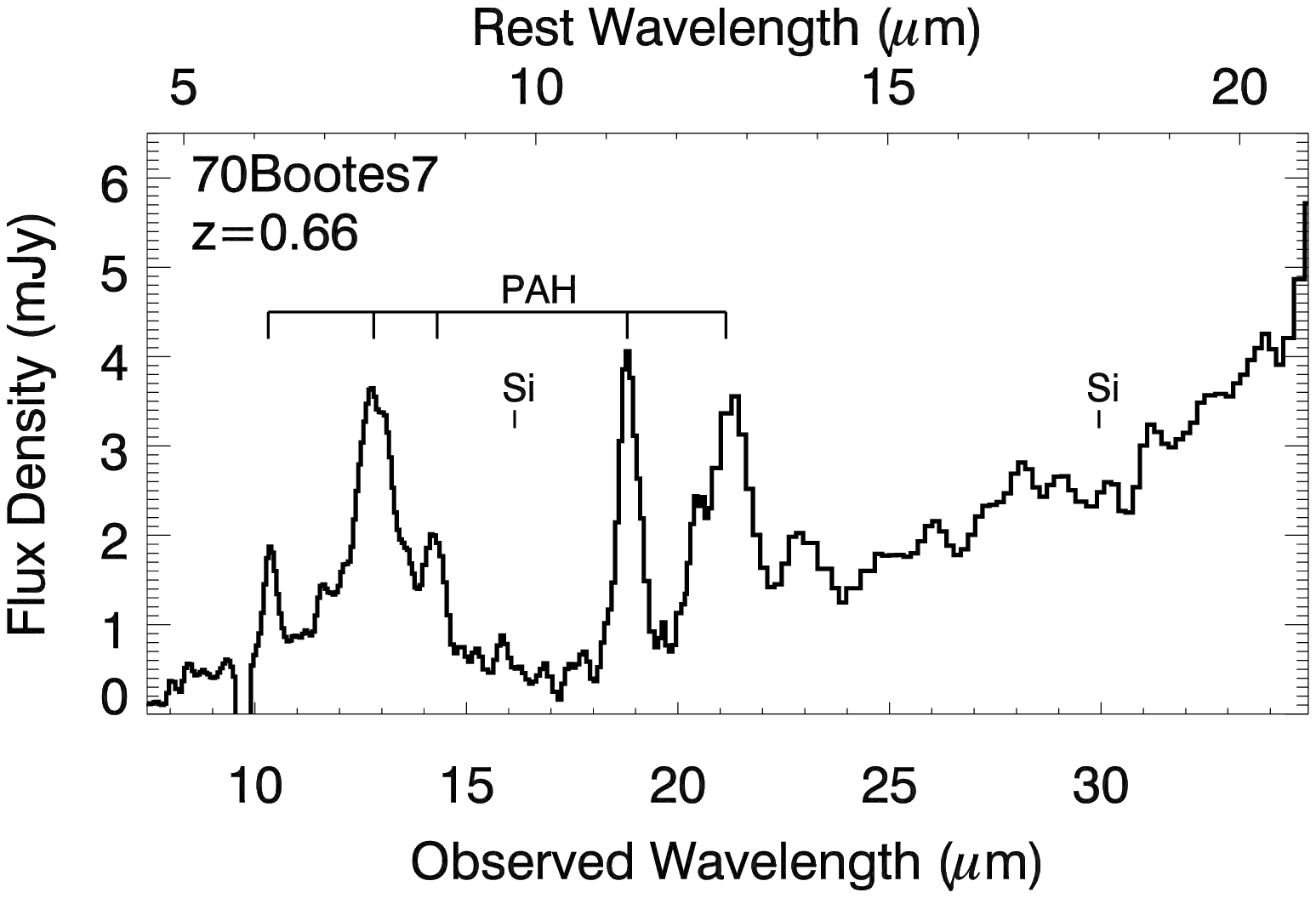}
\includegraphics[height=37mm]{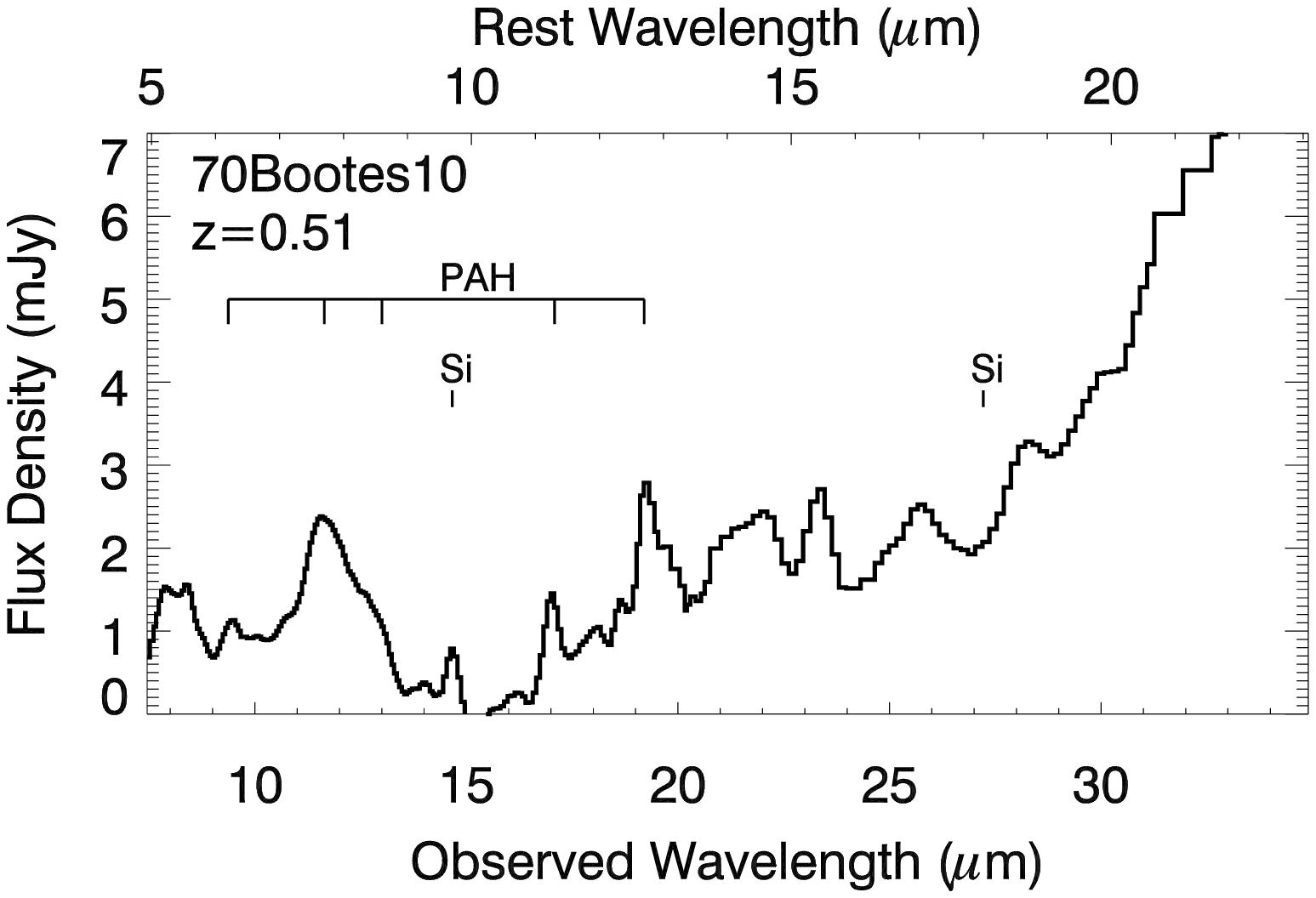}
\includegraphics[height=37mm]{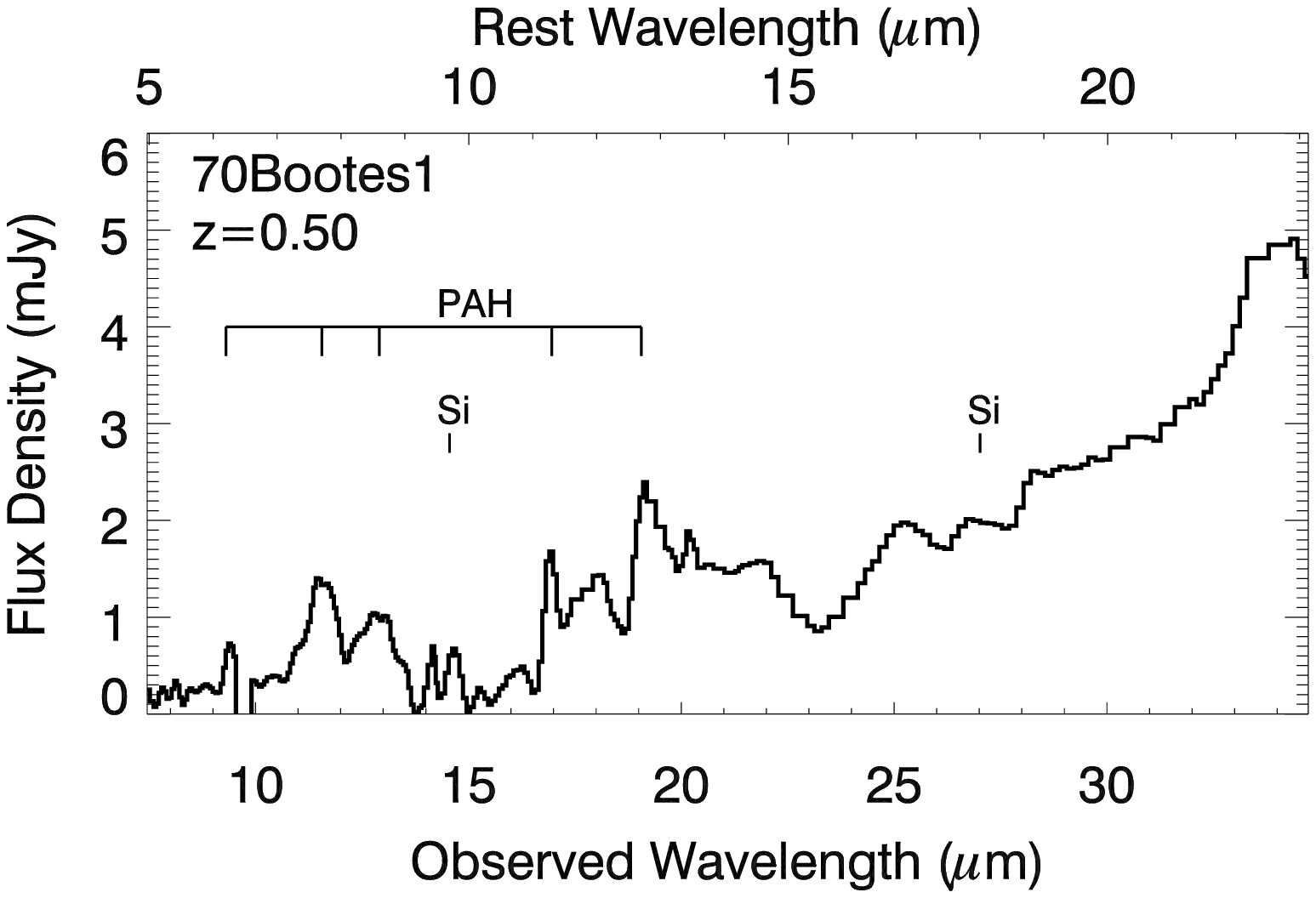}
\includegraphics[height=37mm]{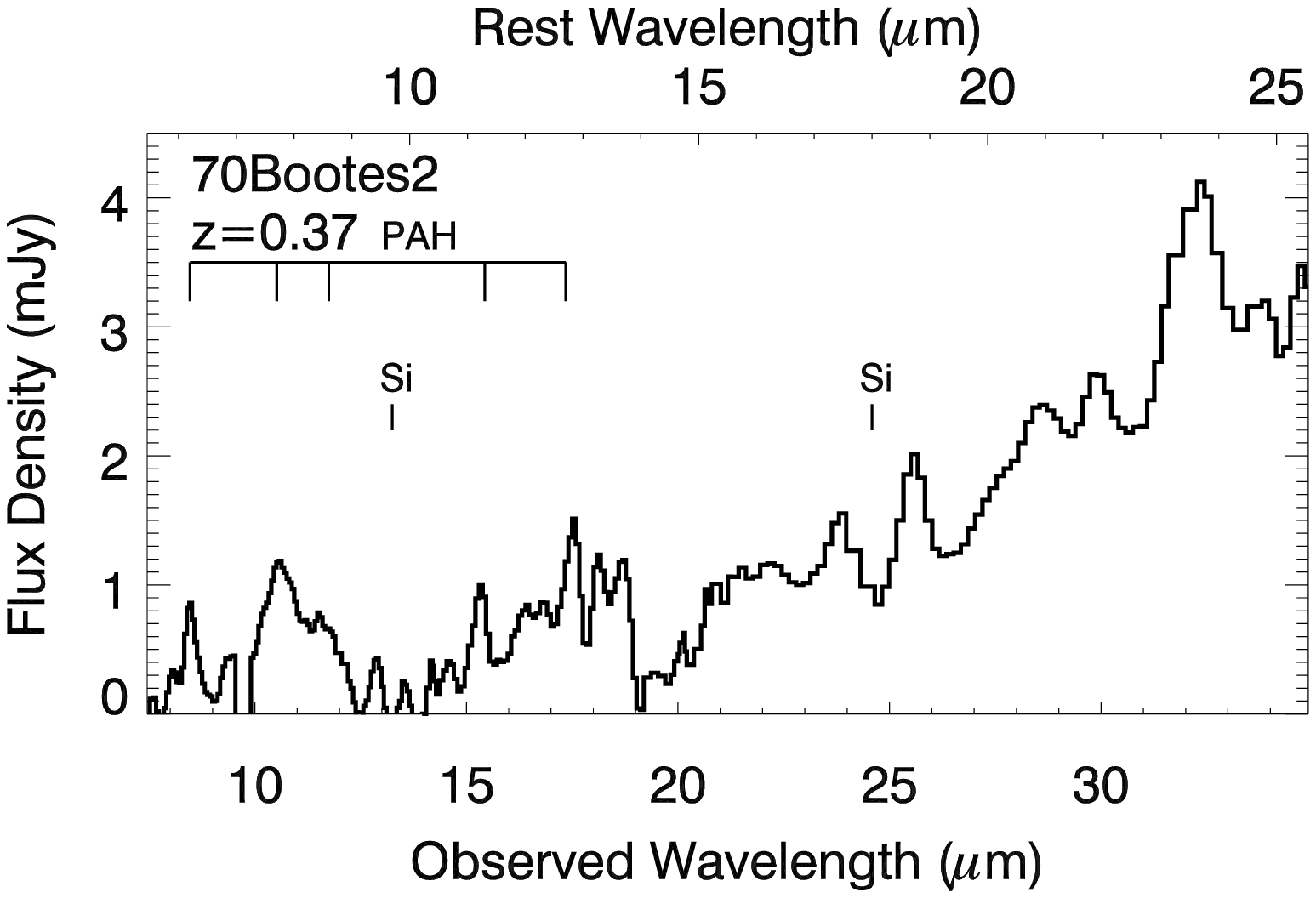}
\includegraphics[height=37mm]{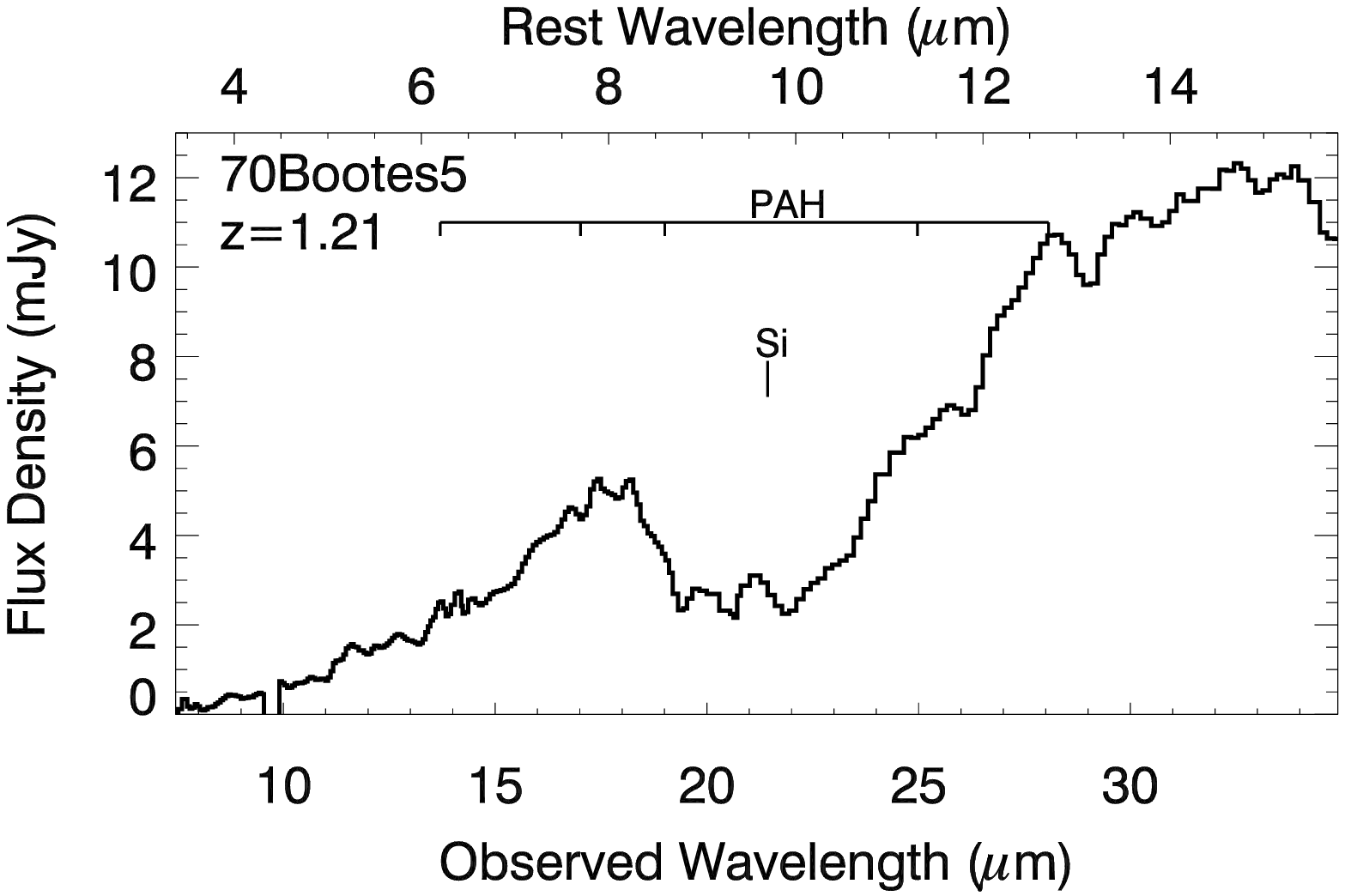}
\includegraphics[height=37mm]{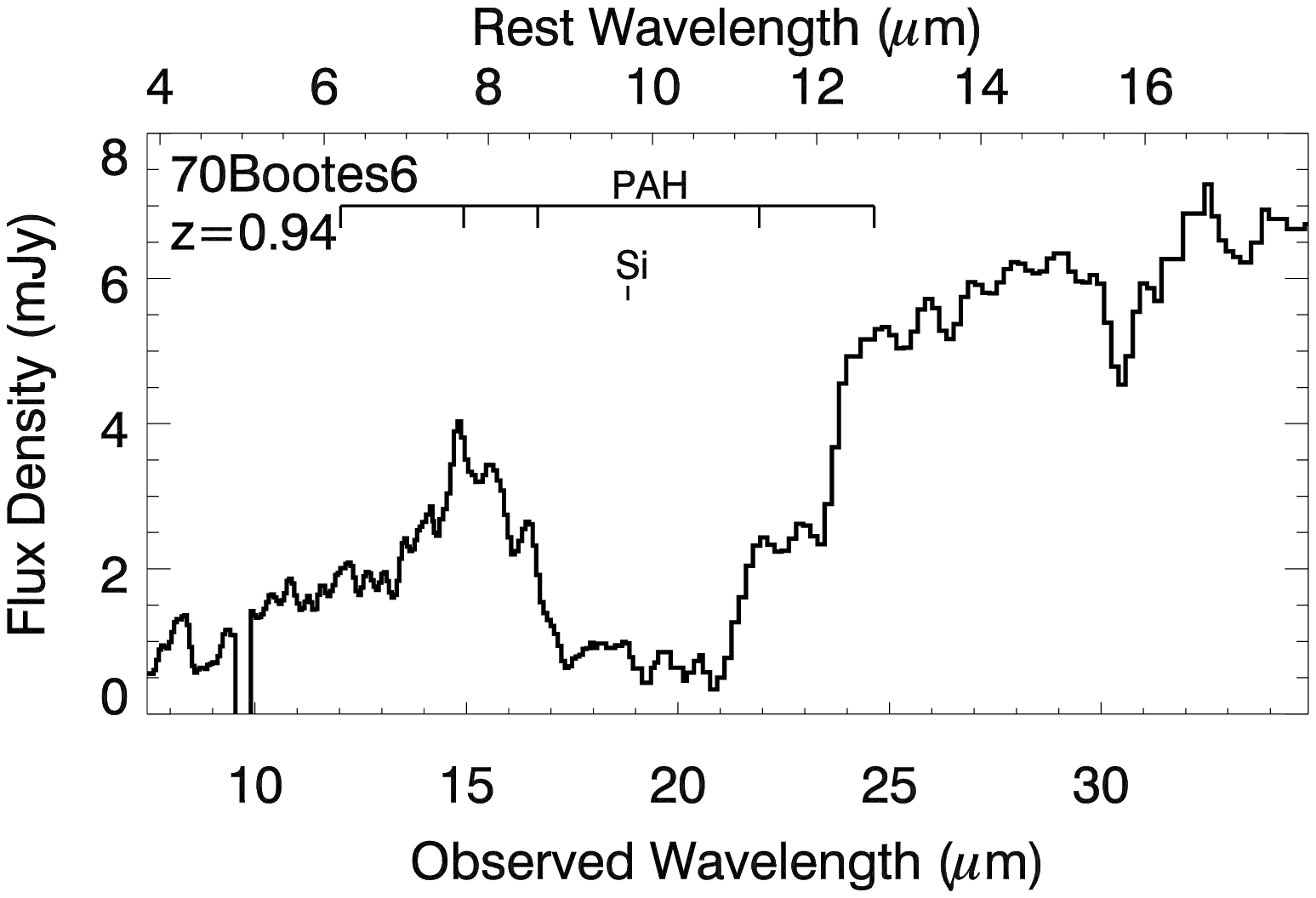}
\includegraphics[height=37mm]{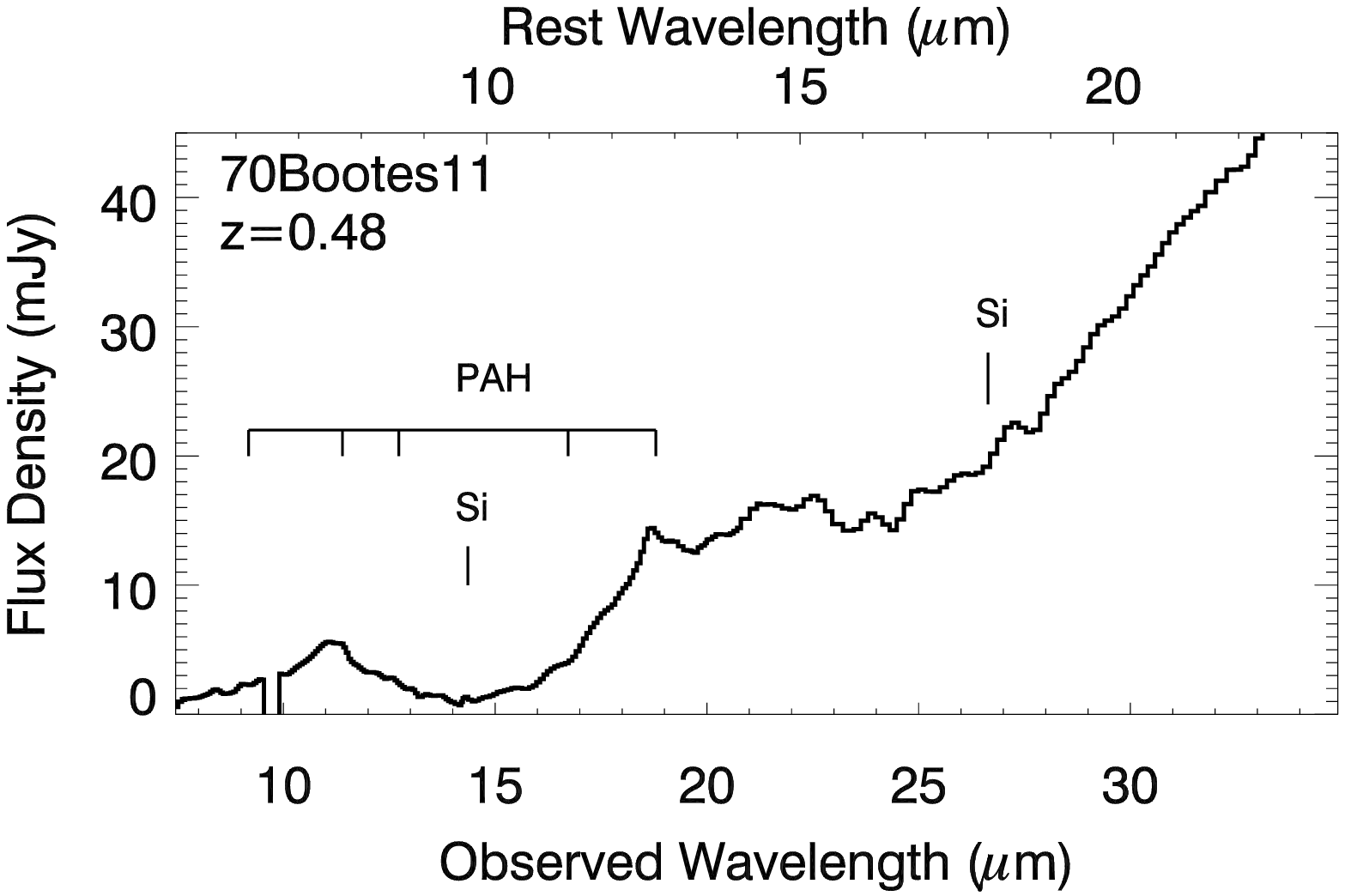}
\includegraphics[height=37mm]{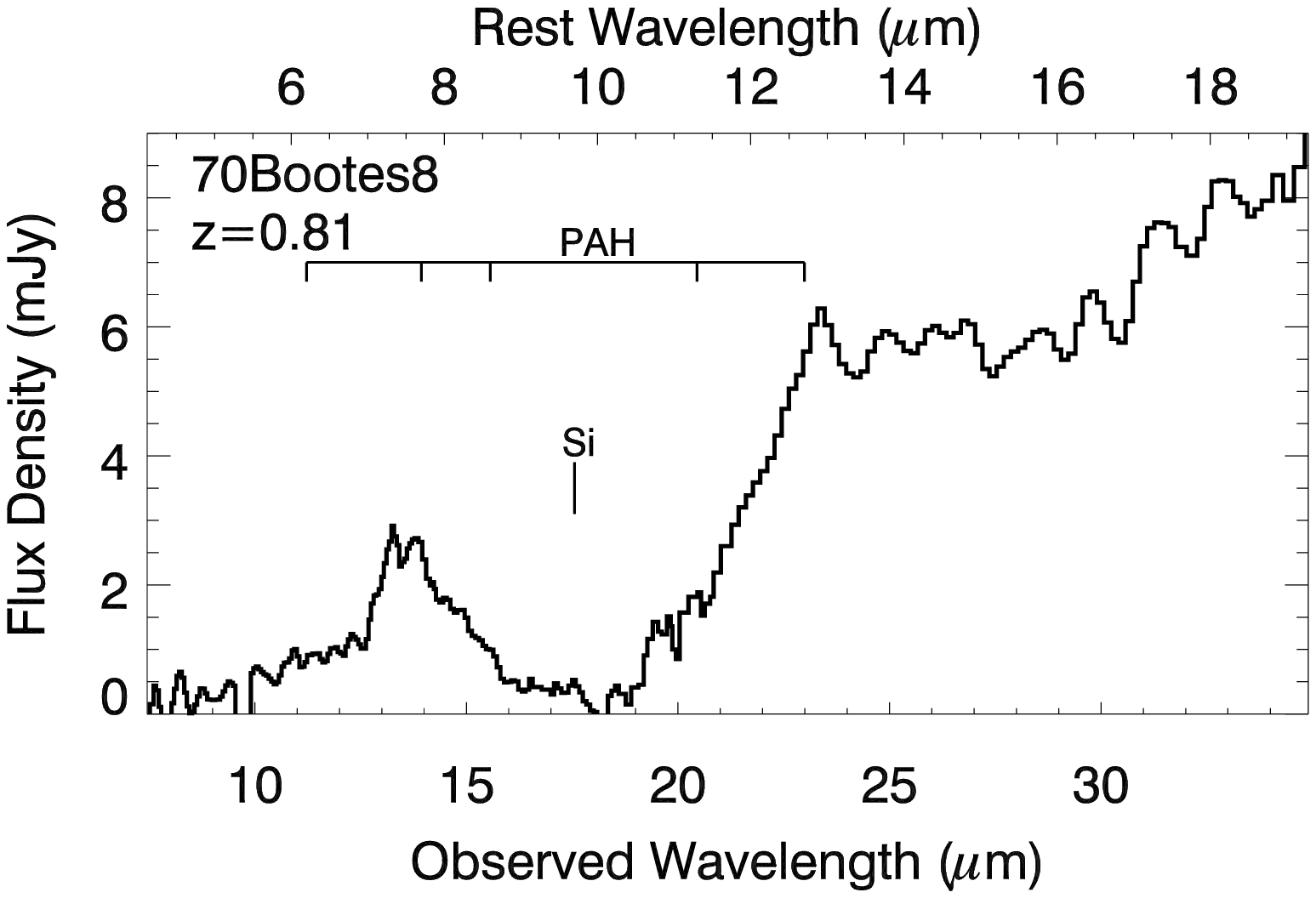}
\includegraphics[height=37mm]{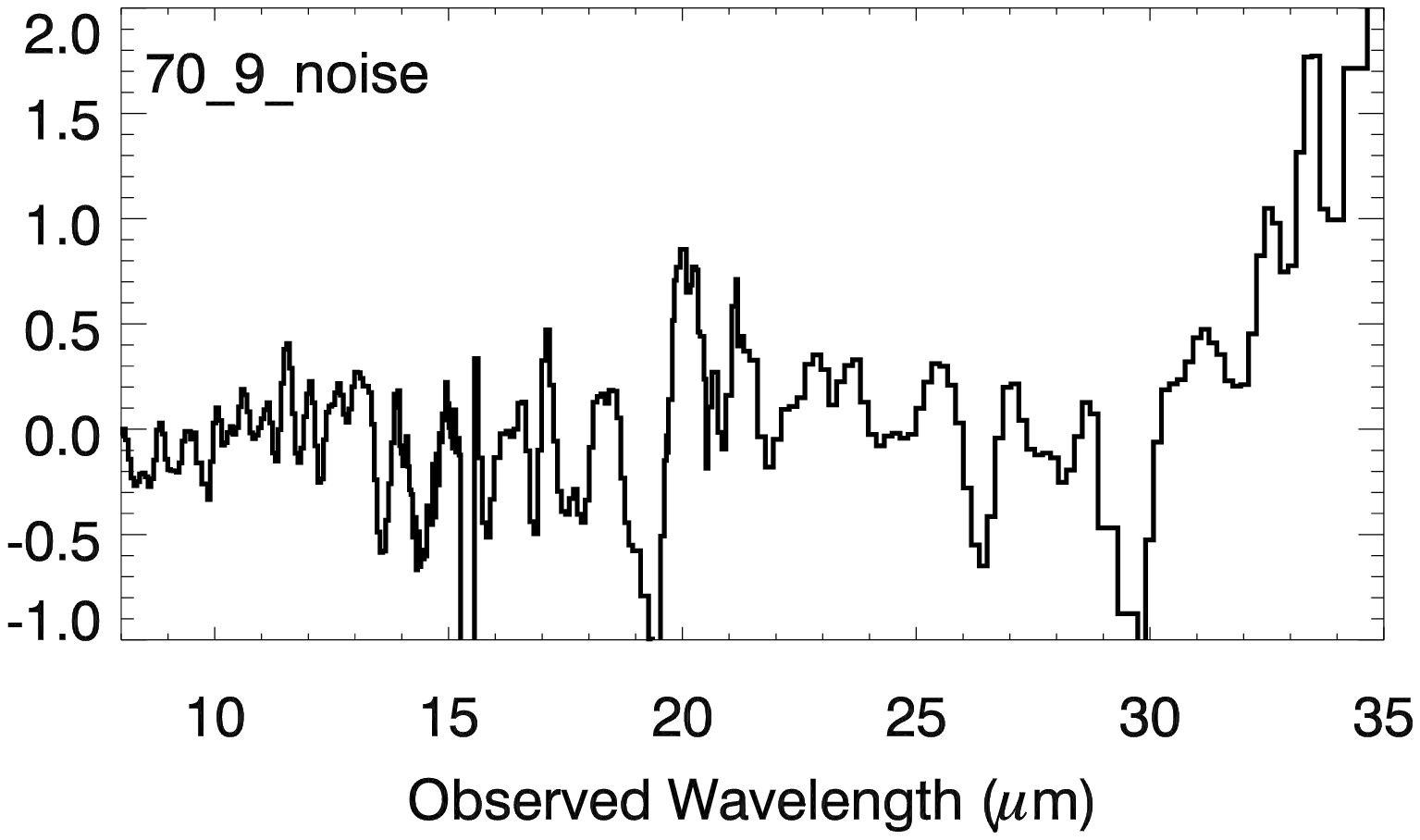}
\end{center}
{\caption[junk]{\label{fig:70spec} IRS spectra of 70 $\rm \mu m$ selected luminous infrared galaxies. The sources are divided into PAH-dominated (top 7) and absorption-dominated (bottom 4) sources. The PAH-dominated sources are ordered by their 7.7$\rm \mu m$ PAH luminosities (from strongest to weakest). The absorption-dominated sources are ordered by their silicate absorption strengths (shallow to deep). The spectra are boxcar smoothed over a resolution element (approximately two pixels). The expected positions of the PAH emission features and silicate absorption features are shown. The measured equivalent widths of the PAH features and silicate absorpion depths are given in Table~\ref{tab:irs}. Also plotted on the bottom right is a typical noise spectrum (for 70Bootes9). }}
\end{figure*}

The IRS spectra of the eleven 70$\rm \mu m$ selected sources are presented in Figure~\ref{fig:70spec}. They are boxcar smoothed over a resolution element (approximately two pixels). Most of the sample (7 of 11) exhibit PAH emission features with measurable 6.2$\rm \mu m$ PAH equivalent widths in their infrared spectra and are hereafter refered to as PAH-dominated sources (denoted as ``sb'' in Table~\ref{tab:irs}). The large PAH equivalent widths are consistent with them being starburst-dominated sources and was anticipated because the large 70$\rm \mu m$ fluxes imply that the spectra are dominated by dust that is cooler than that typically associated with AGN. The remaining four sources exhibit strong 9.7$\rm \mu m$ silicate absorption features, no obvious PAH emission features, and are classified as absorption-dominated sources (denoted as ``abs'' in Table~\ref{tab:irs}). 

For the PAH-dominated sources, redshifts are determined from the strong PAH emission features at rest wavelengths 6.2$\mu$m, 7.7$\mu$m, and 11.3$\mu$m.  With sufficient signal-to-noise ratios, use of these features gives redshifts consistent with optically-derived redshifts to $\pm$0.001 (Houck et al.~2006).  Redshift precision for faint sources depends on the signal-to-noise ratio, so for the spectra with PAH features, we consider the redshifts accurate only to $\pm$0.05. For the absorbed sources, redshifts are determined primarily from the localized maximum in the IRS continuum, or ``hump'', which is blueward of the absorption feature.  In the absorption-dominated sources, this feature is produced by absorption on either side of the feature, and with a possible contribution of 7.7$\mu$m PAH emission in composite sources.  We determine redshifts for absorption-dominated sources by assuming this hump has a rest-frame wavelength of 7.9$\mu$m (see average spectra in \citet{hao07} and \citet{spo07}). Comparison with optical redshifts (e.g., \citealt{bra07}) suggests that they can be uncertain to $\pm$0.1 in $z$. 

\begin{figure}[h]
\begin{center}
\includegraphics[height=55mm]{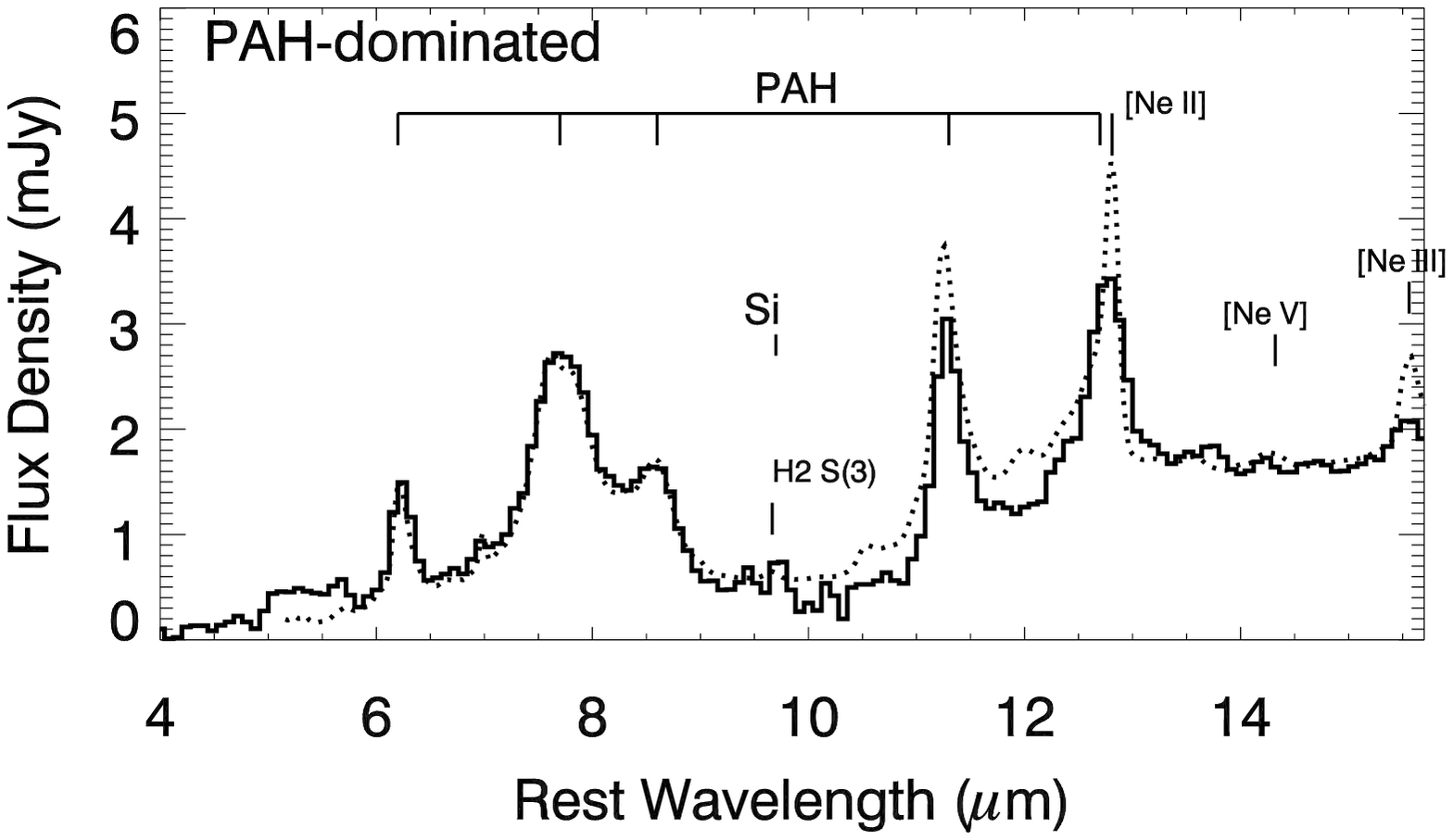}
\includegraphics[height=55mm]{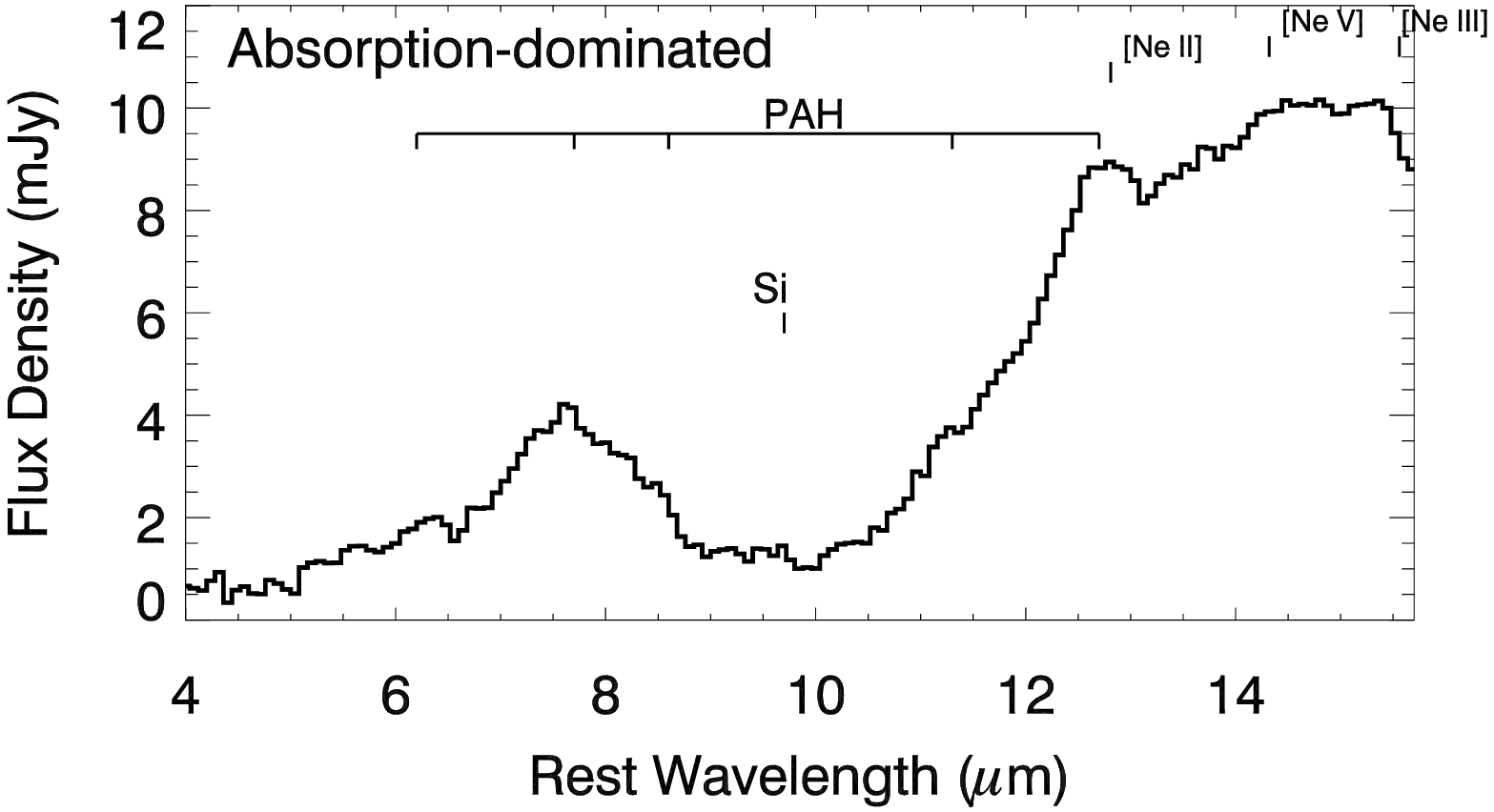}
\end{center}
\caption{\label{fig:xstack} Average IRS spectrum of PAH-dominated sources (top) and absorption-dominated sources (bottom). The expected positions of typically strong emission lines and the silicate absorption line are plotted. The average IRS spectrum of low redshift starburst galaxies from \citet{brl06} is over-plotted for the PAH-dominated spectrum (normalized to the 7.7$\rm \mu m$ peak flux; dotted line).}
\end{figure}

We constructed an average mid-infrared spectra of the 7 sources classified as PAH-dominated and for the 4 sources classified as absorption-dominated. This was achieved by wavelength correcting each individual spectrum to the rest frame ($z=0$), interpolating them to a common wavelength scale of $\approx$0.1$\mu$m/pix, and taking a straight (non-weighted) average at each wavelength position. The resulting average spectra are shown in Figure~\ref{fig:xstack}. We discuss each class in turn. 
\thispagestyle{empty}
\setlength{\voffset}{-24mm}
\setlength{\voffset}{-20mm}

\subsection{PAH-dominated 70 $\rm \mu m$ sources}

The average IRS spectrum of the 7 PAH-dominated sources in Figure~\ref{fig:xstack} is characteristic of classical starbursts. The spectrum is very similar in shape to the average low redshift (and lower luminosity) starburst template presented by \citet{brl06}, although the PAH emission features at 11.3$\rm \mu m$ and 12.7$\rm \mu m$ appear to be weaker in relation to the 7.7$\rm \mu m$ PAH emission feature. The presence of low ionization [Ne II] $\lambda$12.81$\rm \mu m$ and [Ne III] $\lambda$15.56$\rm \mu m$ emission lines and lack of significant high ionization features such as [Ne V] $\lambda$14.32$\rm \mu m$, are also very typical of starbursts. There is a possible detection of the rotational H$_2$ S(3) $\lambda$9.7$\rm \mu m$ emission line in the averaged spectrum. This feature is seen in both high and low resolution IRS spectra of nearby starburst galaxies (e.g., \citealt{brl06}; \citealt{hig06}; \citealt{far07}; \citealt{arm07}) and suggests the presence of warm molecular gas. We measure the rest-frame 6.2$\rm \mu m$ PAH equivalent widths by fitting a single Gaussian and using the continuum adjacent to the feature, between 5.5$\rm \mu m$ and 7$\rm \mu m$. The values shown in Table~\ref{tab:irs} are similar to that of the lower redshift starburst galaxies presented by \citet{brl06} and \citet{des07}.

\begin{figure}[h]
\begin{center}
\includegraphics[height=55mm]{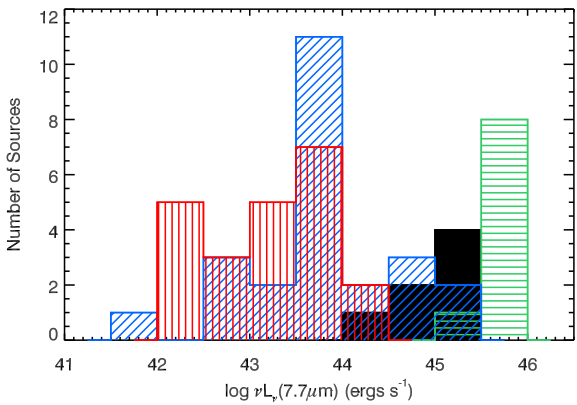}
\end{center}
\caption{\label{fig:l7p7} Histogram showing the 7.7 $\rm \mu m$ PAH luminosities, log($\nu$L$\nu$(7.7$\rm\mu m$)), of the 70 $\rm \mu m$ PAH-dominated sources in our sample (black filled histogram). For comparison, we show the local starbursts from \citealt{brl06} (red vertical stripes), the 10 mJy PAH-dominated sample from Houck et al. 2007 (blue diagonal stripes), and the high redshift starburst sample from \citealt{wee06b} (green horizontal stripes). }
\end{figure}
\begin{figure}[h]
\begin{center}
\includegraphics[height=55mm]{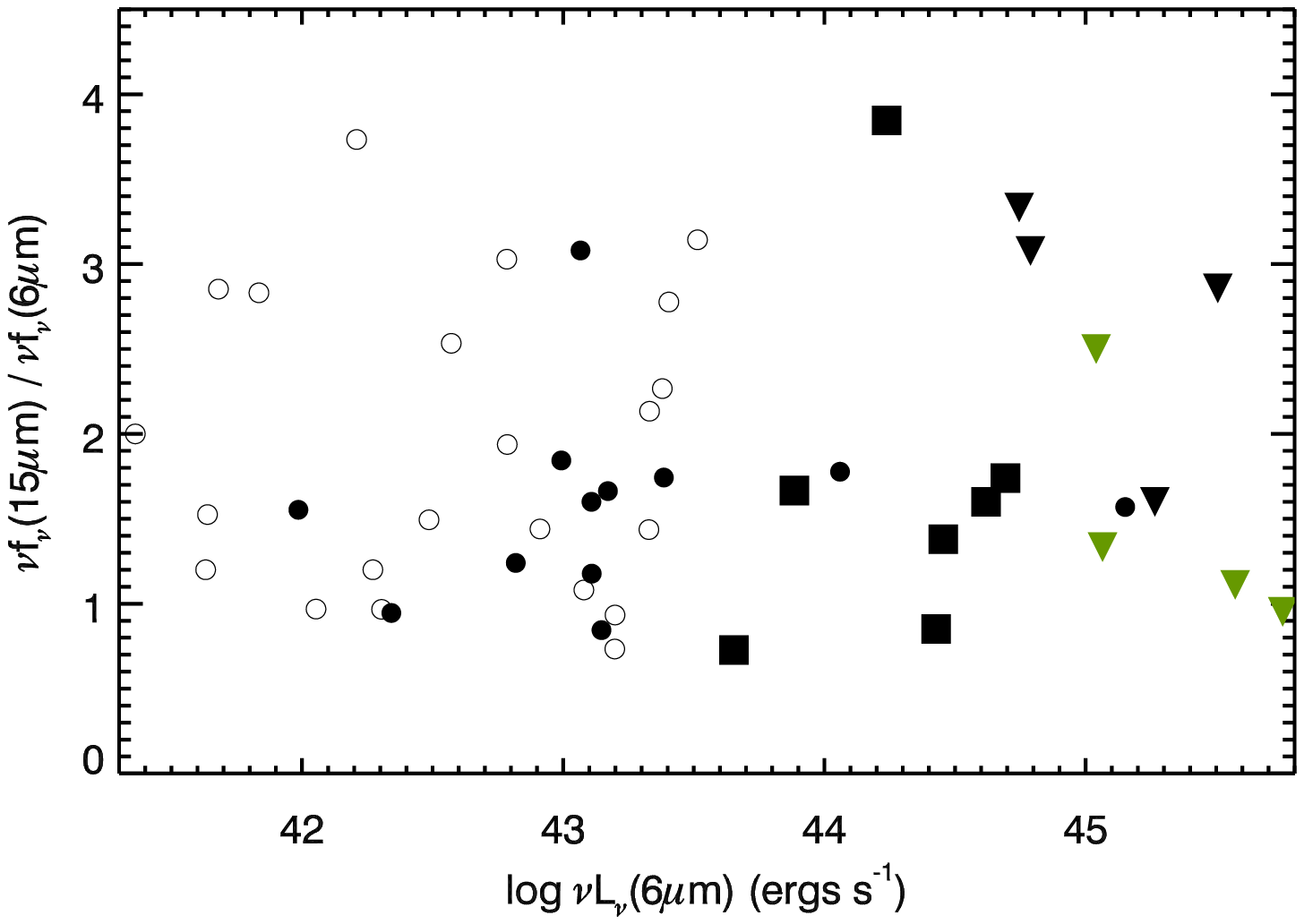}
\includegraphics[height=55mm]{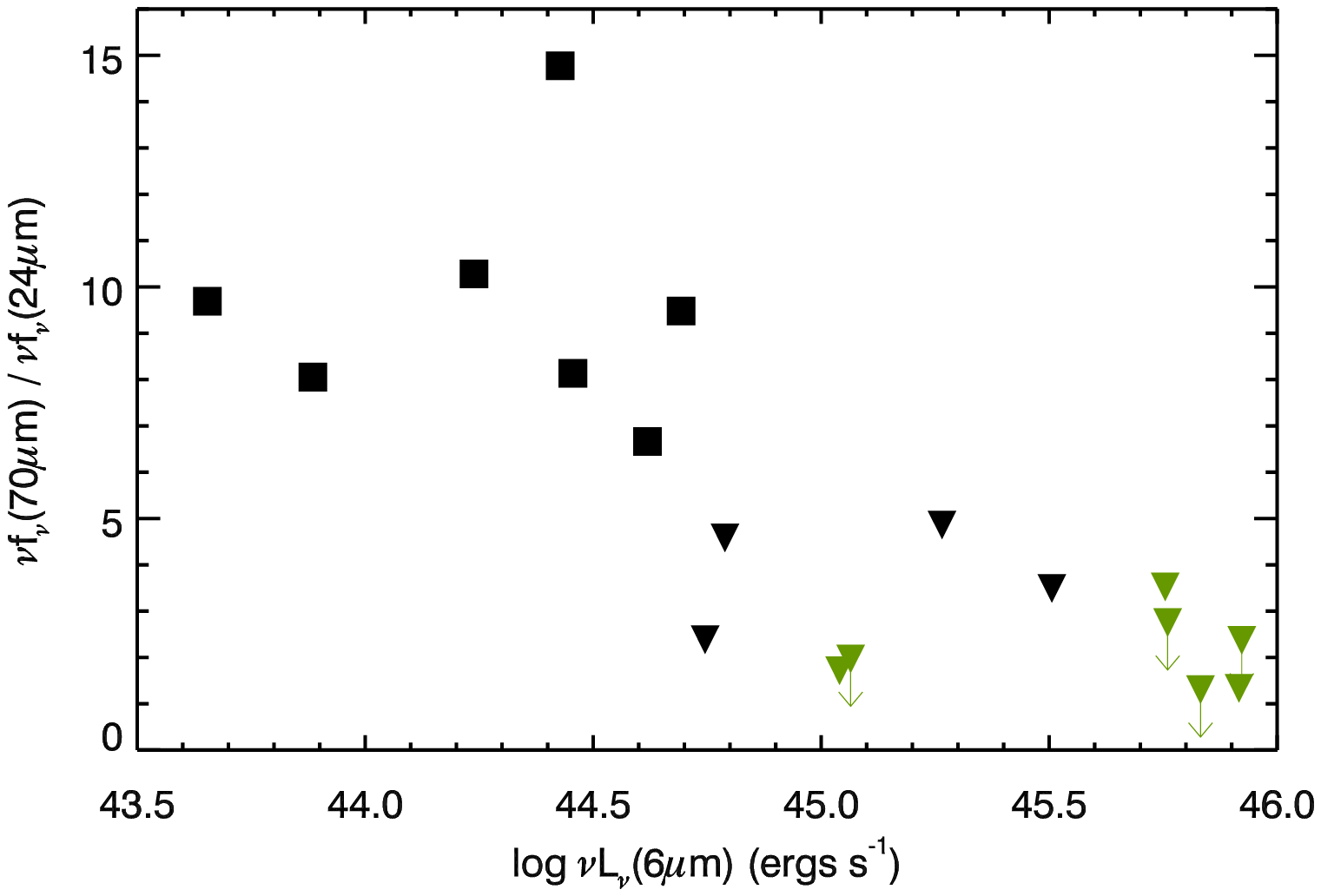}
\end{center}
\caption{\label{fig:l6} Log($\nu$L$_{\nu}$(6$\mu$m)) as a function of rest-frame $\nu$f$_{\nu}$(15$\mu$m)/$\nu$f$_{\nu}$(6$\mu$m) ratio (top) and observed frame $\nu$f$_{\nu}$(70$\mu$m)/$\nu$f$_{\nu}$(24$\mu$m) ratio (bottom). PAH-dominated sources are represented by large black squares; absorption-dominated sources are represented by large black triangles; small filled circles represent PAH emission sources from the Bo\"otes 10 mJy sample (Houck et al. 2007); small empty circles represent local starbursts from \citet{brl06}; large green triangles represent X-ray-selected AGN from Brand et al.~(2007).}
\end{figure}

These sources are interesting because of their extreme mid-infrared luminosities and red colors. The median continuum luminosity is $\nu$L$_{\nu}$(6$\mu$m)=4 $\times$ 10$^{44}$ ergs s$^{-1}$ (1 $\times$ 10$^{11}$ L$_{\odot}$).  For comparison, the median luminosity for starbursts in the Bo\"otes f$_{24}>$10 mJy sample of Houck et al. (2007) is 1 $\times$ 10$^{43}$ ergs s$^{-1}$ (3 $\times$ 10$^{9}$ L$_\odot$) and for the prototype starburst NGC 7714 of \citet{brl06} is 5 $\times$ 10$^{42}$ ergs s$^{-1}$ (1 $\times$ 10$^9$ L$_\odot$). In Figure~\ref{fig:l7p7} we compare the PAH luminosities, log($\nu$L$_{\nu}$(7.7$\mu$m)), of our sample to other starbursts. The mean PAH luminosity of the PAH-dominated sources is orders of magnitude larger than that of the local starbursts from \citet{brl06}. The least luminous sources in our sample are comparable to the most luminous starbursts in the local Universe. The luminosities are slightly lower than that of the high redshift starbursts from \citet{wee06b} which were chosen from the existence of a stellar photospheric peak in their near-IR SEDs. If we use the conversion in Houck et al. (2007) between the 7.7 $\rm \mu m$ PAH luminosity and star formation rate, we find that our most luminous source has a star-formation rate of $\sim$720 M$_\odot$ yr$^{-1}$. This is only $\sim$50\% lower than the upper limit on the H$\alpha$ estimated star formation rates in local bright galaxies \citep{ken98}. Table~\ref{tab:irs} shows that these sources have larger infrared luminosities (as estimated from their 24$\rm \mu m$, 70$\rm \mu m$, and 160$\rm \mu m$ luminosities) than is estimated from their 7.7 $\rm \mu m$ PAH luminosities. Their infrared luminosities are among the largest known for starforming galaxies.  
We measure the rest-frame f$_{\nu}$(6$\mu$m) and f$_{\nu}$(15$\mu$m) flux densities from the IRS spectra. f$_{\nu}$(6$\mu$m) is measured just shortward of 6$\rm \mu m$ to avoid the 6.2$\rm \mu m$ PAH feature and provide a measurement of the hot dust continuum. In Figure~\ref{fig:l6}, we show the distribution of 6$\mu$m luminosities ($\nu$L$_{6\rm \mu m}$) as a function of continuum slope as measured by the rest-frame flux density ratio, $\nu$f$_{\nu}$(15$\mu$m)/$\nu$f$_{\nu}$(6$\mu$m) and observed flux density ratio, $\nu$f$_{\nu}$(70$\mu$m)/$\nu$f$_{\nu}$(24$\mu$m). The PAH-dominated sources have a lower mean $\nu$f$_{\nu}$(15$\mu$m)/$\nu$f$_{\nu}$(6$\mu$m) than that of the lower luminosity starbursts of \citet{brl06}. Although this could suggest a contribution from an AGN to the infrared emission (since the hot dust results in a shallower spectral slope), a larger sample size is needed to show this with any confidence. 

\subsection{Absorption-dominated 70 $\rm \mu m$ sources}

The four absorption-dominated 70$\rm \mu m$ sources have IRS spectra which are generally more characteristic of AGN: they have no obvious PAH emission features (6.2$\rm \mu m$ PAH equivalent width $<$ 0.03$\rm \mu m$) but deep silicate absorption features. They have 6$\rm \mu m$ luminosities that are larger than that of the PAH-dominated sources and similar to that of the X-ray selected AGN in Brand et al.~(2007). The infrared luminosities are also very large (L$_{IR}\sim~0.3-1.7\times10^{13}~\rm L_{\odot}$). 

We measure the silicate strength S$_{10}$,
\begin{equation}
S_{10}=ln\frac{f_{obs}(10 \rm \mu m)}{f_{cont}(10 \rm \mu m)},
\end{equation}
 using the method of \citet{spo07} where $f_{obs}(10 \rm \mu m)$ is the observed flux density at the peak of the 10$\rm \mu m$ feature and $f_{cont}(10 \rm \mu m)$ is the continuum flux at the peak wavelength, extrapolated from the continuum to either side. The silicate absorption depths are S$_{10}$= $-$1.0, $-$1.9, $-$3.6, and $-$2.2 for 70Bootes5, 70Bootes6, 70Bootes8, and 70Bootes11 respectively. These correspond to very heavy absorption, similar to the median for ULIRGs in \citet{hao07} and \citet{spo07} (S$_{10}$= $-$1.6) and places them among the most heavily absorbed sources known (the most extreme example known has S$_{10}$= $-$4.0; \citealt{spo06}). \citet{hao07} find that the silicate strength correlates with the infrared slope at high mid-IR wavelengths (14-30$\rm \mu m$) but not at lower mid-IR wavelengths (5-14 $\rm \mu m$), supporting the idea that the silicate feature arises in the cooler dust (although note that no correlation between the f$_\nu$(60$\rm \mu m$)/f$_\nu$(25$\rm \mu m$) flux density ratio and silicate strength is seen by \citet{ima07} in their sample of low redshift ULIRGs). Give that our selection criteria preferentially picks sources which are bright at 70 $\rm \mu m$, the large silicate strengths for the absorption-dominated sources are consistent with the findings of \citet{hao07}. 

Are these sources powered by AGN or starburst activity? The individual spectra show no significant PAH-emission features (although there is a hint of the 7.7, 8.6, and 12.7$\rm \mu m$ PAH features in the averaged spectrum in Figure~\ref{fig:xstack}), so they appear most similar to class 3A (little to no PAH emission, strong silicate absorption) in the scheme of \citet{spo07}. The IRS spectra are similar to that of 2 sources with deep silicate absorption strengths in a sample of 87 local Seyfert galaxies presented by \citet{buc06}. \citet{far07} suggest that sources with very deep silicate absorption are likely to be AGN-dominated sources. The four sources have the largest 6$\rm \mu m$ rest-frame luminosities in the sample and large infrared luminosities, again suggesting AGN activity. We observe no significant [Ne V] emission line that we might expect for AGN-dominated sources. However, given that [Ne V] often has a low equivalent width and that there are only 4 IRS spectra with low signal-to-noise at the observed wavelength of [Ne V] (see noise spectrum in Figure~\ref{fig:70spec}), this is not strong evidence against a large contribution from AGN emission.

Because AGN-dominated galaxies typically have shallower mid-IR slopes than starburst-dominated galaxies (e.g., \citealt{bra06b}), measuring flux density ratios at wavelengths free of strong absorption or emission lines may help in determining their primary power source. For heavily absorbed sources, estimating the slope at large rest-frame mid-infrared wavelengths is important because dust may absorb near-IR photons and re-emit them at longer wavelengths, steepening the spectral slope. Figure~\ref{fig:l6} shows that the absorption-dominated sources have observed-frame $\nu$f$_{\nu}$(70$\mu$m)/$\nu$f$_{\nu}$(24$\mu$m) flux density ratios that are smaller than the PAH-dominated sources and closer to that of the X-ray selected AGN presented in Brand et al.~(2007), suggesting that they have shallower infrared spectral slopes indicative of AGN-dominated sources. We note that one must be cautious in interpreting these results since the flux densities are in the observed frame and the rest-frame flux density ratio may vary with redshift. Although we can directly measure rest-frame flux density ratios at shorter wavelengths, these wavelengths are more likely to be affected by absorption. The f$_{\nu}$(15$\mu$m)/f$_{\nu}$(6$\mu$m) flux density ratios of absorption-dominated sources are generally larger than that of the PAH-dominated sources. We suggest that strong dust absorption in the absorption-dominated sources results in a steeper infrared slope at these shorter infrared wavelengths, despite the fact that they have small $\nu$f$_{\nu}$(70$\mu$m)/$\nu$f$_{\nu}$(24$\mu$m) flux density ratios and are likely to be dominated by AGN. 

We conclude that these are examples of very obscured sources, which are likely to be powered by AGN. However, we cannot rule out the possibility of them being powered by deeply embedded starbursts. If the mid-IR emission of these sources is dominated by AGN, this implies that $\approx$36\% of sources selected to have luminous 70 $\rm \mu m$ emission are AGN-dominated. Perhaps the 70 $\rm \mu m$ emission, which traces cooler dust, is also powered largely by AGN (see e.g., \citealt{ima07}). In this cases, large 70 $\rm \mu m$ luminosities be generated because the AGN are so heavily absorbed by dust that a large fraction of the hot dust emission is absorbed and re-emitted at longer wavelengths. Alternatively, if a large fraction of the mid-IR emission originates from star-formation activity, this implies that star-formation activity as traced by mid-IR signatures can be hidden by obscuration and/or low level AGN activity. Follow-up multi-wavelength observations (optical spectroscopy, deep X-ray and radio imaging) are clearly needed to fully determine the nature of these rare sources.

\section{Infrared and infrared-to-optical colors of 70 $\rm \mu m$ selected sources}

\begin{figure}[h]
\begin{center}
\includegraphics[height=55mm]{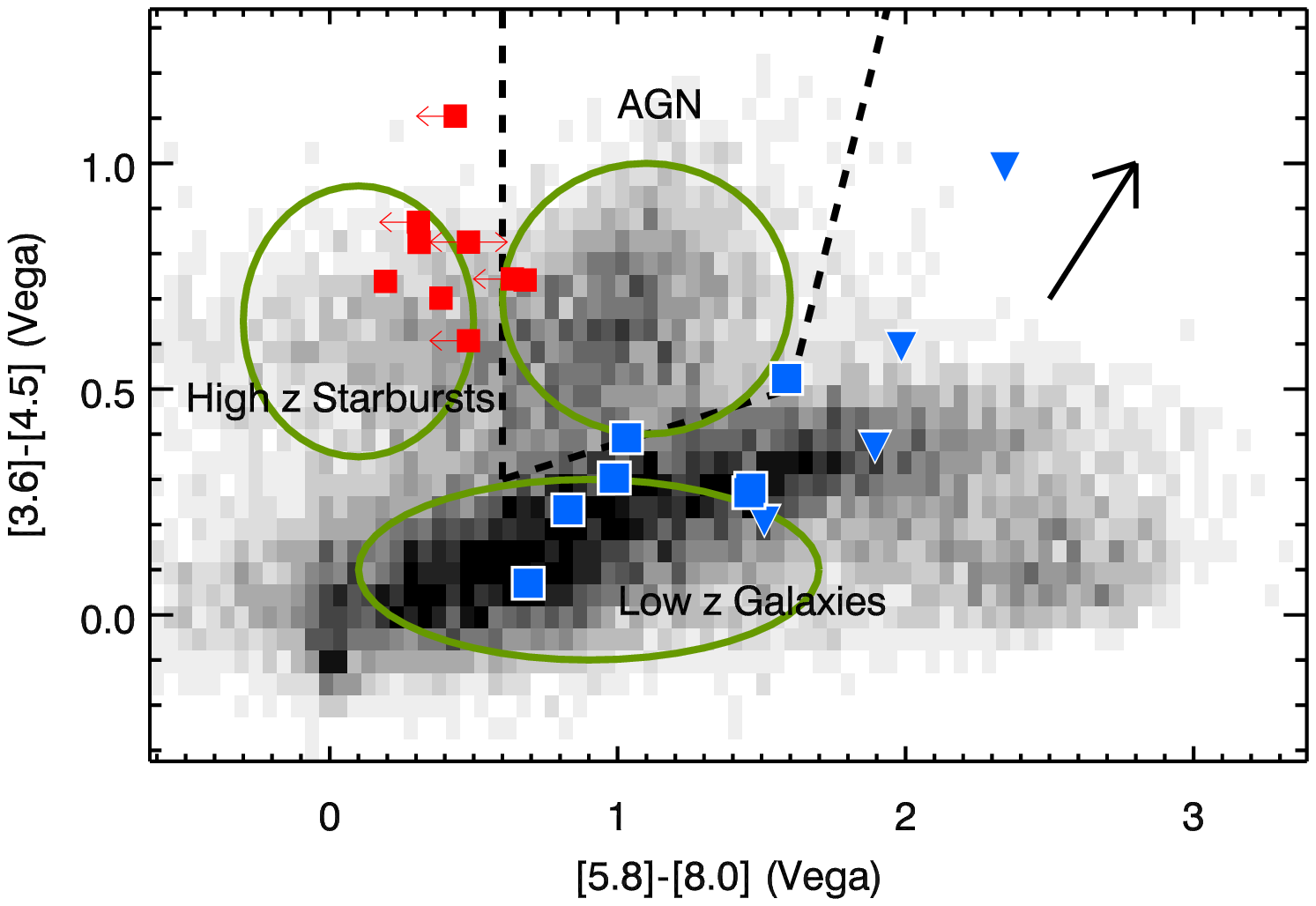}
\includegraphics[height=55mm]{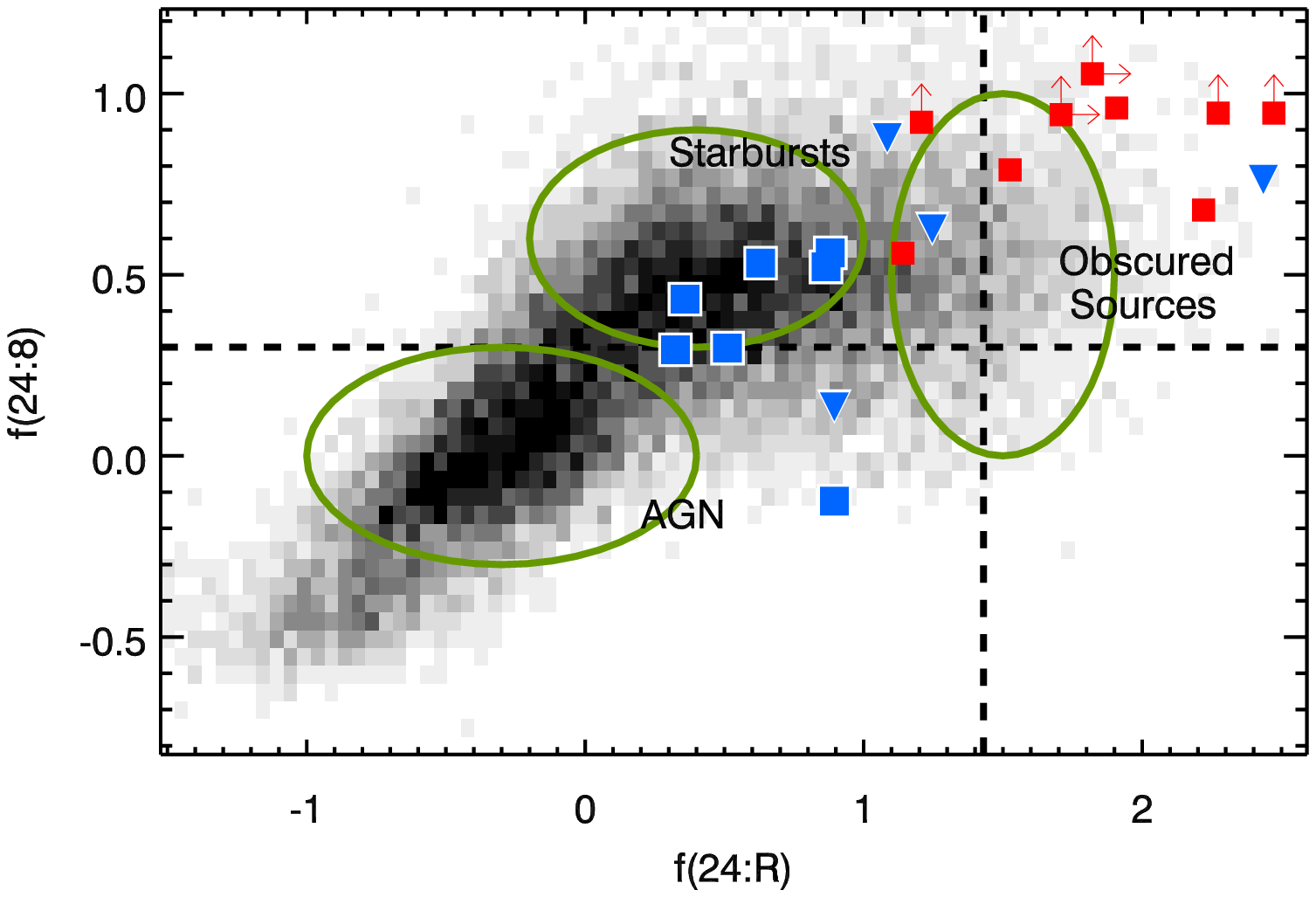}
\end{center}
{\caption[junk]{\label{fig:irac} IRAC color-color diagram (top) and 24-to-8 $\rm \mu m$ vs. 24-to-0.7 $\rm \mu m$ color-color diagram (bottom) for 70 $\rm \mu m$ selected IRS sources from Bo\"otes (large blue symbols) and starburst-dominated sources from \citet{wee06b} (smaller red squares). The Bo\"otes sources are split into silicate absorption-dominated sources (large blue downwards triangles) and PAH-dominated sources (large blue squares). We show the distribution of all $\approx$10,000 MIPS sources with f$_{24}>$0.5 mJy in greyscale. The rough regions expected to be inhabited by different populations are also shown. The black dashed line in the top figure is that proposed by \citet{ste05} to empirically separate AGN from Galactic stars and normal galaxies. The arrow shows the reddening curve estimated from the R$_V$=3.1 dust model of \citet{dra03} for a $z$=1 source. In the bottom figure, the vertical dashed line shows the $R-[24]>$14 (log($\nu$f$_\nu$(24)/$\nu$f$_\nu$(R))$>$1.43) criteria used to selected the powerful obscured sources presented in \citet{hou05} and \citet{wee06} and the horizontal dashed line shows the 24-to-8$\rm \mu m$ color criteria used by \citet{bra06b} to roughly divide steeper spectrum starburst sources from shallow spectrum AGN-dominated sources at $z>$0.6.}}
\end{figure}

To help provide further insight as to the nature of the PAH-dominated and absorption-dominated sources, we show the IRAC color-color diagram for our IRS sources in Figure~\ref{fig:irac}a. For both the PAH-dominated and absorption-dominated sources, the  colors are consistent with the expected positions of starburst-dominated relatively low redshift galaxies (see also \citealt{yan04}; \citealt{saj05}). Although we have concluded in previous sections that the absorption-dominated sources are probably largely powered by AGN emission, they do not fall into the ``AGN wedge'' of \citet{ste05}. This is likely due to very heavy obscuration of the AGN. The IRAC colors of the absorption-dominated sources are consistent with their emission coming predominantly from heavily reddened star-formation processes (see reddening arrow estimated from the R$_V$=3.1 dust model of \citet{dra03}). The IRAC colors are redder than the PAH-dominated sources, consistent with them being more dusty and obscured than the PAH-dominated sources.

We also plot the IRAC colors of the starburst-dominated sources selected by \citet{wee06b} to have optical and near-IR SEDs showing a luminosity peak from stellar photospheric emission at $1.0<z<1.9$. These sources have (by design) red [3.6]$-$[4.5] colors but blue [5.8]$-$[8.0] colors and confirm that sources in this region tend to be high redshift starburst-dominated sources. This region becomes more heavily populated at lower 24 $\rm \mu m$ fluxes, consistent with the population becoming more dominated by starbursts sources at fainter f$_{24}$ as found by \citet{bra06b}. 
\thispagestyle{empty}
To investigate the effects of obscuration, we show the 24-to-8 $\rm \mu m$ vs. 24-to-0.7 $\rm \mu m$ color-color diagram for our sources in Figure~\ref{fig:irac}b,(see also \citet{yan04} for the location of different populations in this color-color space). The 24-to-0.7 $\rm \mu m$ color is a good indicator of obscuration whereas the 24-to-8 $\rm \mu m$ color can be used as a crude measure of the spectral slope and hence the dust temperature distribution and whether the source is AGN- or starburst-dominated (\citealt{bra06b} find that AGN-dominated sources tend to have f(24:8)$<$0.3 and starburst-dominated sources tend to have f(24:8)$>$0.3; see also \citealt{yan04}). The PAH-dominated 70 $\rm \mu m$ selected sources have lower f(24:R) ratios than the absorption-dominated sources. Because PAH emission features can fall into the 8$\rm \mu m$ IRAC wave-band for $z<$0.6, the 24-to-8 $\rm \mu m$ color is not a good indicator for most of these sources. The absorption-dominated 70$\rm \mu m$ sources have higher f(24:R) ratios than the PAH-dominated sources and steep spectral slopes, again consistent with them being more obscured. The starburst sources from \citet{wee06b} have high f(24:R) ratios and f(24:8) ratios consistent with them being higher redshift, obscured starburst-dominated sources. 

\section{Multi-wavelength SEDs}

\begin{figure}[h]
\begin{center}
\includegraphics[height=55mm]{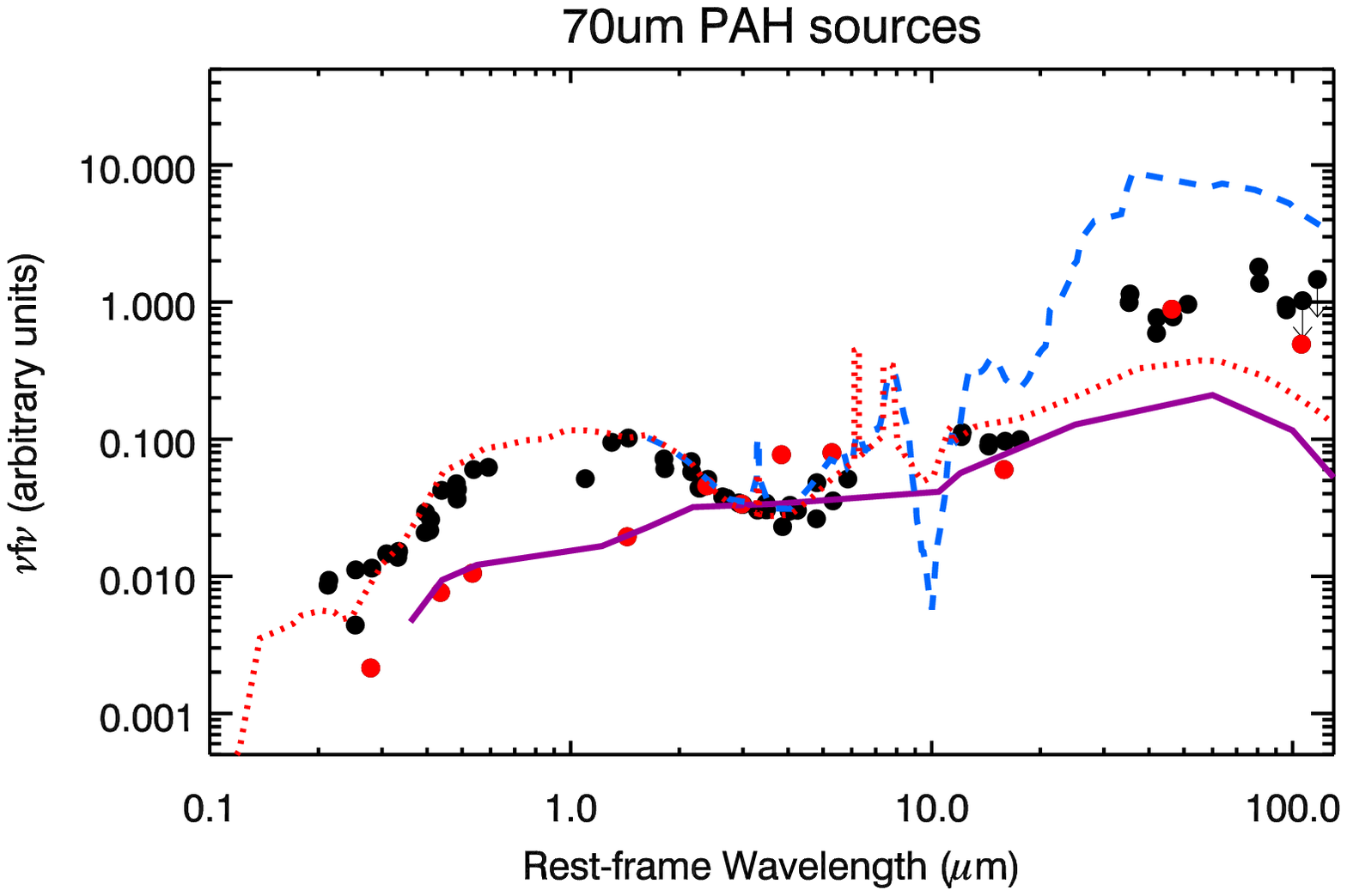}
\includegraphics[height=55mm]{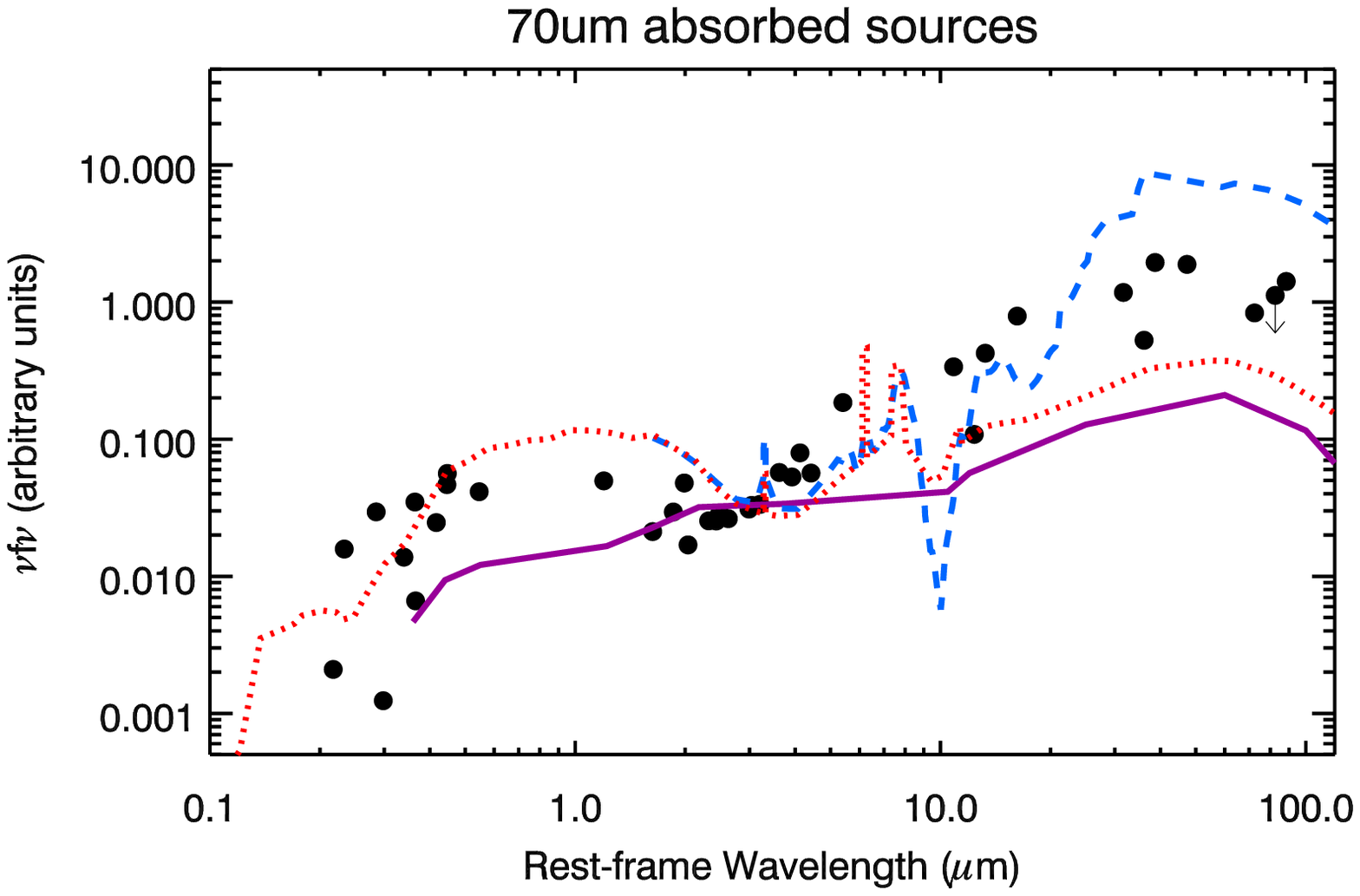}
\end{center}
\caption{\label{fig:sed} Optical to far-IR SEDs for PAH-dominated sources (top) and absorption-dominated sources (bottom). The individual spectra are scaled to have the same 3$\rm \mu m$ flux density. Also plotted are SED templates of Mrk231 (solid purple line), Arp220 (dashed blue line; \citealt{sil98}), and M82 (dotted red line; \citealt{sil98}). For the PAH-dominated spectra, most SEDs are similar to M82 below $\sim$15$\rm \mu m$. 70Bootes10 is not well fit by M82 and its SED points are shown in red.}
\end{figure}

In Figure~\ref{fig:sed}, we show the rest-frame multi-wavelength ($B_w$, $R$, $I$, $K$, 3.6$\rm \mu m$, 4.5$\rm \mu m$, 5.8$\rm \mu m$, 8$\rm \mu m$, 24$\rm \mu m$, 70$\rm \mu m$, 160$\rm \mu m$) SEDs for the PAH-dominated and absorption-dominated sources. We also show the SEDs of M 82, Arp 220, and Mrk 231. M 82 is the nearest example of a star-bursting galaxy and is often considered the prototype of the starburst phenomena (e.g., \citealt{rie80}). Arp 220 is a starburst-dominated ULIRG with strong dust extinction (e.g., \citealt{soi84}; \citealt{stu96}). Mrk 231 is a local obscured AGN with a luminous circumnuclear starburst (e.g., \citealt{sol92}; \citealt{wee05}). The individual SEDs are normalized to the same flux density at 3$\rm \mu m$. All but one of the PAH-dominated sources are well fit by M 82 blue-wards of 15$\rm \mu m$, consistent with them being starburst-dominated sources. 70Bootes10 (shown by the red points) is better fit by Mrk 231, suggesting that there is a large contribution of AGN emission in this source. 70Bootes10 is perhaps one of the most likely of the PAH-dominated sources to host an energetically important AGN, since it has the lowest 6.2$\rm \mu m$ PAH equivalent width of all the PAH-dominated sources. The far-IR flux densities are between that of M 82 and Arp 220, suggesting that the PAH-dominated sources have a large relative cool to hot dust content that is between these two local sources. The absorption-dominated sources have a larger variation in their SED shapes. None of the sources exhibit a strong near-IR ``bump'' which is characteristic of starburst galaxies. The far-IR slopes are steeper than that of Mrk 231, suggesting that these sources have a larger cool to hot dust ratio than Mrk 231. This may result from our selection criteria which requires the galaxies to be very luminous at 70$\rm \mu m$. 

\section{Conclusions}

We have presented $Spitzer$ IRS mid-infrared spectra of a small sample of 11 sources from the $Spitzer$ MIPS survey of the Bo\"otes field using selection criteria based on 70$\mu$m detections (f$_{70}>$30 mJy) and faint optical magnitudes ($R>20$). All the sources lie in the redshift range $0.3<z<1.3$, implying very large mid- and far-infrared luminosities. The IRS spectra of the eleven galaxies show either PAH emission features (7/11) or deep silicate absorption (4/11) features. The IRS spectra of the seven PAH-dominated sources are typical of classical starbursts, but with extremely large mid-infrared luminosities. This sample contains sources with very high PAH luminosities, implying star-formation rates of up to 700 M$_{\odot}$ yr$^{-1}$. The PAH-dominated sources tend to have lower $\nu$f$_{\nu}$(15$\mu$m)/$\nu$f$_{\nu}$(6$\mu$m) flux density ratios than that of lower redshift starburst galaxies with lower infrared luminosities. A larger sample is needed to show whether this is consistent with a contribution to the f$_{\nu}$(6$\mu$m) luminosity from hot dust associated with an AGN in more luminous sources.

The four absorption-dominated sources have very deep silicate absorption features that are larger than the median of that of local ULIRGs and their large $\nu$f$_{\nu}$(15$\mu$m)/$\nu$f$_{\nu}$(6$\mu$m) flux density ratios suggest that their mid-infrared spectral slope is steep. This suggests that the energy source is embedded in large volumes of cool dust. The very red [5.8]-[8.0] and relatively red [3.6]-[4.5] colors also suggest a large amount of dust along the line of sight to these sources. The deep silicate absorption, small $\nu$f$_{\nu}$(70$\mu$m)/$\nu$f$_{\nu}$(24$\mu$m) flux density ratio (which is less affected by dust obscuration than estimates from lower infrared wavelengths), and multi-wavelength SEDs suggests that they are likely dominated by AGN, although we cannot rule out that they are powered by very obscured starbursts. Follow-up multi-wavelength observations such as optical spectroscopy, and deep X-ray and radio observations may help to confirm their power source. 

We thank our colleagues on the NDWFS, MIPS, IRS, AGES, and IRAC teams. KB is supported by the Giacconi fellowship at STScI. Support for this work by the IRS GTO team at Cornell University was provided by NASA through Contract Number 1257184 issued by JPL/Caltech. This research is partially supported by the National Optical Astronomy Observatory which is operated by the Association of Universities for Research in Astronomy, Inc. (AURA) under a cooperative agreement with the National Science Foundation. This work is based on observations made with the {\it Spitzer Space Telescope}, which is operated by the Jet Propulsion Laboratory, California Institute of Technology under a contract with NASA. The {\it Spitzer / MIPS} and IRAC surveys of the Bo\"otes region were obtained using GTO time provided by the {\it Spitzer} Infrared Spectrograph Team (James Houck, P.I.), M. Rieke, and the IRAC Team (G. Fazio, P.I.). IRAC is supported in part through contract 960541 issued by JPL. The IRS was a collaborative venture between Cornell University and Ball Aerospace Corporation funded by NASA through the Jet Propulsion Laboratory and the Ames Research Center.


\end{document}